%% file: paper-aaai25-arxiv.tex
\documentclass[letterpaper]{article} %
\usepackage{aaai25}  %
\usepackage{times}  %
\usepackage{helvet}  %
\usepackage{courier}  %
\usepackage[hyphens]{url}  %
\usepackage{graphicx} %
\usepackage{natbib}  %
\usepackage{caption} %
\usepackage{algorithm}

\usepackage{newfloat}
\usepackage{listings}
\DeclareCaptionStyle{ruled}{labelfont=normalfont,labelsep=colon,strut=off} %
\floatstyle{ruled}
\newfloat{listing}{tb}{lst}{}
\floatname{listing}{Listing}
\usepackage[switch, modulo]{lineno}

\title{Parallel Greedy Best-First Search with a Bound on the Number of Expansions Relative to Sequential Search}
\author{
    Takumi Shimoda\textsuperscript{\rm 1},
    Alex Fukunaga\textsuperscript{\rm 2}\\
}
\affiliations{
    
    Graduate School of Arts and Sciences\\
    The University of Tokyo\\
    takumi35shimoda@yahoo.co.jp\textsuperscript{\rm 1}, fukunaga@idea.c.u-tokyo.ac.jp\textsuperscript{\rm 2}
}

\usepackage{bibentry}
\usepackage{comment}
\usepackage{xspace}
\usepackage{bm}

\usepackage{subcaption}
\usepackage{multirow}
\usepackage{algorithm}
\usepackage{amsmath}
\usepackage{amsfonts}
\usepackage{amsthm}
\usepackage[noend]{algpseudocode}
\usepackage{booktabs} %
\usepackage{xr} 
\usepackage{amssymb} %
\usepackage{varwidth}%
\theoremstyle{definition}
\newtheorem{def.}{Definition}
\newtheorem{th.}{Theorem}

\newtheorem{lem.}{Lemma}
\newtheorem{cor.}{Corollary}
\newtheorem{prop.}{Proposition}
\newtheorem{obs.}{Observation}

\newcommand{\pddl}[1]{\textsf{\small #1}}

\newcommand{\open}{\mathit{Open}}
\newcommand{\closed}{\mathit{Closed}}

\newcommand{\parent}{\mathit{parent}}

\newcommand{\NULL}{\mathit{NULL}}
\newcommand{\Succ}{\mathit{succ}}
\newcommand{\hwm}{\mathit{hwm}}

\newcommand{\BTS}{\mathit{BTS}}

\newif\ifwastar
\wastarfalse
\newcommand{\astar}{$\mathit{A^*}$}

\newcommand{\Deferred}{\mathit{Deferred}}   %
\newcommand{\Unevaluated}{\mathit{Unevaluated}}   %
\newcommand{\OBATS}{OBAT$_S$\xspace}
\newcommand{\PDUHF}{PUHF3$_S$\xspace}
\newcommand{\KPDGBFS}{KPGBFS$_S$\xspace}

\NewDocumentCommand{\rot}{O{90} O{1em} m}{\makebox[#2][l]{\rotatebox{#1}{#3}}}

\begin{document}

\maketitle

\begin{abstract}
  Parallelization of non-admissible search algorithms such as GBFS poses a challenge because straightforward parallelization can result in search behavior which significantly deviates from  sequential search.
  Previous work proposed PUHF, a parallel search algorithm which is constrained to only expand states that can be expanded by some tie-breaking strategy for GBFS.
  We show that despite this constraint, the number of states expanded by PUHF is not bounded by a constant multiple of the number of states expanded by sequential GBFS with the worst-case tie-breaking strategy.
  We propose and experimentally evaluate One Bench At a Time (OBAT), a parallel greedy search which guarantees that the number of states expanded is within a constant factor of the number of states expanded by sequential GBFS with some tie-breaking policy.
\end{abstract}

\section{Introduction}

Best-first search (BFS) algorithms such as \astar \cite{HartNR68} and GBFS \cite{DoranM66} are a key component of
planning and scheduling systems.
In order to fully exploit the increasing number of cores on modern CPUs,
it is necessary to parallelize these BFS algorithms.
Successful approaches to parallelization have been developed for 
admissible BFS algorithms such as \astar \cite{BurnsLRZ10,KishimotoFB13,PhillipsLK14}. %

Parallelization of GBFS %
has turned out to be more challenging.
Parallel portfolios which execute completely independent instances of GBFS (with a different tie-breaking on each thread), e.g., \cite{KuroiwaF2020} are a simple approach.
However, this approach does not exploit the potential for explicit cooperation among threads.

A natural approach to implement a more ``cooperative'' multi-threaded search is to use shared data structures ($\open$, $\closed$), similar to parallel \astar.
It is easy to implement a state expansion {\it policy} which seems similar to sequential GBFS, e.g., ``expand a state with the best heuristic value in the (shared) $\open$ list''.
However the resulting state expansion {\it behavior} of such a parallel GBFS can deviate drastically from that of sequential GBFS.
This poses two issues:
(1) the relationship between the states expanded by parallel vs. sequential GBFS is nontrivial to analyze theoretically, and it is difficult to answer basic questions such as: ``what is the worst-case performance of this parallel GBFS relative to sequential GBFS?''
(2) For some problems, the empirical performance of parallel GBFS is much worse than sequential GBFS, resulting in orders of magnitude of slowdown on some IPC benchmarks \cite{KuroiwaF2019}, and in general,
the number of states expanded by
simple approaches to parallel GBFS with shared $\open$ and/or $\closed$ lists is not bounded by a constant factor relative to sequential GBFS
\cite{KuroiwaF2020}.

Previous work proposed PUHF \cite{KuroiwaF2020}, a parallel GBFS algorithm which is constrained to only expand states in the Bench Transition System (BTS), the set of states that can be expanded by some tie-breaking strategy for GBFS \cite{HeusnerKH17}.
The BTS provides a natural theoretical constraint on the states expanded by an algorithm which ``behaves similarly to sequential GBFS'', somewhat analogous to the (much stronger) constraint that a parallel \astar must expand all states with evaluation value less than the optimal.
However, although PUHF guarantees that only states in the BTS are expanded, it does not guarantee a bound on the number of states expanded compared to sequential GBFS, i.e., PUHF provides a qualitative guarantee about search behavior, but it does not provide any performance guarantee.

Thus, previous parallel GBFS algorithms with shared $\open$/$\closed$ lists pose an {\it unbounded risk} -- on any given instance, parallel GBFS may outperform sequential GBFS,
  or it may perform much worse, with no known general bound on how badly search performance can degrade.
In this paper, we directly address the problem of establishing a bound on the number of states expanded by parallel GBFS relative to sequential GBFS.
While PUHF constrained expansion to only states which could be expanded by sequential GBFS with some tie-breaking strategy,
we propose a new algorithm which further constrains the search so that
it explores only one bench at a time. This allows us to bound the number of state expansions by parallel search to be no worse than the number of states expanded by sequential search with some tie-breaking policy plus a small constant.

The main contributions of this paper are:
(1) We propose One Bench At a Time (OBAT), the first parallel search algorithm with shared $\open$/$\closed$ lists with a guaranteed bound on the number of expansions relative to sequential GBFS.
(2) We show that this theoretical guarantee does not require sacrificing practical performance -- OBAT, combined with SGE (a method that parallelizes both state generation and evaluation) \cite{ShimodaF24}, performs comparably to baselines which have no such guarantee.

The rest of the paper is structured as follows.
After a review of preliminaries and background (Sec. \ref{sec:preliminaries}), 
we first show that the number of states expanded by PUHF is not bounded by a constant multiple of the number of states expanded by GBFS with the worst-case tie-breaking strategy (Sec. \ref{sec:puhf-pathology}).
We then propose One Bench At a Time (OBAT), which guarantees that the number of states expanded is within a constant factor of the number expanded by sequential GBFS with some tie-breaking policy (Sec. \ref{sec:bepuhf}-\ref{sec:bepuhf-analysis}).
We also propose \OBATS, which significantly improves the state evaluation rate by using SGE (Sec. \ref{sec:obats}).
We experimentally show that although the number of states expanded by OBAT
is competitive with previous parallel algorithms, the state evaluation rate is significantly lower. However, we show that \OBATS is competitive with previous methods (Sec. \ref{sec:bepuhf-experiment}).

\section{Preliminaries and Background}
\label{sec:preliminaries}

\subsubsection{State Space Topology}
State space topologies are defined following \citet{HeusnerKH18}.
 
\begin{def.}
  A {\it state space} is a 4-tuple $\mathcal{S} = \langle S, \Succ,
  s_{init}, S_{goal} \rangle$,
  where $S$ is a finite set of states,
  $\Succ : S \rightarrow 2^S$ is the successor function,
  $s_{init} \in S$ is the initial state,
  and $S_{goal} \subseteq S$ is the set of goal states.
  If $s' \in \Succ(s)$, we say that $s'$ is a successor of $s$ and that $s \rightarrow s'$ is a (state) transition.
  $\forall s \in S_{goal}, \Succ(s) = \emptyset$.
  A {\it heuristic} for $\mathcal{S}$ is a function $h : S \rightarrow \mathbb{R}$
  and $\forall s \in S_{goal}, h(s) = 0$. %
  A {\it state space topology} is a pair $\langle \mathcal{S}, h \rangle$, where $\mathcal{S}$ is a state space.
\end{def.}

We call a sequence of states $\langle s_0, ..., s_n \rangle$  a {\it path} from $s_0$ to $s_n$,
and denote the set of paths from $s$ to $s'$ as $P(s, s')$.
$s_i$ is the $i$ th state in a path $p$ and $|p|$ is the length of $p$.
A {\it solution} of a state space topology is a path $p$ from $s_{init}$ to a goal state.
We assume at least one goal state is reachable from $s_{init}$,
 and  %
$\forall s \in S, s \notin \Succ(s)$.

\subsubsection{Best-First Search}
Best-First Search (BFS) is a class of search algorithms that use an evaluation function $f : S \rightarrow \mathbb{R}$ and a tie-breaking strategy $\tau$.
BFS searches states in the order of evaluation function values ($f$-values).
States with the same $f$-value are prioritized by $\tau$.
In Greedy Best-First Search (GBFS; \citeauthor{DoranM66} \citeyear{DoranM66}), $f(s) = h(s)$. %

\subsubsection {K-Parallel GBFS (KPGBFS)}
\begin{algorithm}[htb]
  \begin{algorithmic}[1]
    \scriptsize
    \State $\open \leftarrow \{ s_{init} \}, \closed \leftarrow \{ s_{init} \}; \forall i, s_i \leftarrow \NULL$
    \For{$i \leftarrow 0,...,k - 1$ in parallel} \Comment{$k$ is the number of threads}
      \Loop
        \While{$s_i=\NULL$}
        \State lock($\open$)
        \If{$\forall j, s_j = \NULL$}
          \If{$\open = \emptyset$} unlock($\open$); \Return $NULL$ \EndIf
        \Else
          \State $s_i \leftarrow top(\open)$; $\open \leftarrow \open \setminus \{ s_i \}$
        \EndIf
        \State unlock($\open$)
        \EndWhile
        \If{$s_i \in S_{goal}$} \Return $s_i$ \EndIf
        \State{lock($\open$), lock($\closed$)}
        \For{$s_i' \in \Succ(s_i)$}
          \If{$s_i' \notin \closed$}
            \State $\closed \leftarrow \closed \cup \{ s_i' \}$
            \State  $\open \leftarrow \open \cup \{ s_i' \}$;
          \EndIf
        \EndFor
        \State{unlock($\open$), unlock($\closed$)}
        \State $s_i \leftarrow \NULL$
      \EndLoop
    \EndFor
    \caption{K-Parallel GBFS (KPGBFS)}
    \label{alg:KPGBFS}
  \end{algorithmic}
\end{algorithm}
K-Parallel BFS \cite{VidalBH10} is a straightforward, baseline parallelization of BFS.
All $k$ threads share a single $\open$ and $\closed$.
Each thread locks $\open$ to remove
a state $s$ with the lowest $f$-value in $\open$, locks $\closed$ to
check duplicates and add $\Succ(s)$ to $\closed$, and locks $\open$ to
add $\Succ(s)$ to Open.
KPGBFS is KPBFS with $f(s)=h(s)$.
Algorithm \ref{alg:KPGBFS} shows the pseudocode for KPGBFS.

\subsubsection{T-bounded and T-pathological}
Kuroiwa and Fukunaga (\citeyear{KuroiwaF2020}) proposed the following in order to classify the quantitative behavior of parallel BSF algorithms.

\begin{def.} 
  Given a state space topology $\mathcal{T}$,
  a search algorithm $A$ is {\it $t$-bounded} relative to search algorithm $B$ on $\mathcal{T}$
  iff $A$ performs no more than $t$-times as many expansions as $B$.
  $A$ is {\it $t$-pathological} relative to search algorithm $B$ on $\mathcal{T}$
  iff $A$ is not $t$-bounded relative to $B$ on $\mathcal{T}$.

  $A$ is $t$-bounded relative to $B$
  iff $A$ is $t$-bounded relative to $B$ for all state space topologies.
  $A$ is $t$-bounded relative to $B$ on state space topologies with the property $P$
  iff $A$ is $t$-bounded relative to $B$ for all state space topologies with $P$.

  $A$ is {\it pathological} relative to $B$
  iff for all
$t > 0$ there exists a state space topology $\mathcal{T}$,
  such that $A$ is $t$-pathological relative to $B$ on $\mathcal{T}$.
  $A$ is pathological relative to $B$ on state space topologies with property $P$
  iff for all
$t > 0$ there exists a state space topology $\mathcal{T}$ with property $P$
  such that $A$ is $t$-pathological relative to $B$ on $\mathcal{T}$.
\end{def.}

Kuroiwa and Fukunaga (\citeyear{KuroiwaF2020}) showed that straightforward parallelizations of GBFS with shared $\open$ and/or $\closed$, including KPGBFS, are pathological.

\subsubsection{Bench Transition Systems}
Heusner et al. (\citeyear{HeusnerKH17}) defined progress states and bench transition systems in order to characterize the behavior of GBFS, building upon the definition of high-water marks by Wilt and Ruml (\citeyear{WiltR14}).

\begin{def.}
  Let $\langle \mathcal{S}, h \rangle$ be a state space topology with states $S$
  and $P(s) = \{ p \in P(s, s') \mid s' \in S_{goal} \}$.
  The high-water mark of $s \in S$ is
  \[ \hwm(s) := \left\{
           \begin{array}{ll}
             \min_{p \in P(s)}(\max_{s' \in p} h(s')) & \textnormal{if $P(s) \neq \emptyset$} \\
             \infty & \textnormal{otherwise}
           \end{array}
         \right.
  \]
  The high-water mark of a set of states  $S' \subseteq S$ is defined as

  \[\hwm(S') := \min_{s \in S'} \hwm(s)
    \]
    \label{def:hwm}
\end{def.}

\begin{def.}
  A state $s$ of a state space topology $\langle \mathcal{S}, h \rangle$ is a {\it progress state}
  iff $\hwm(s) > \hwm(\Succ(s))$.
  \label{def:progress}
\end{def.}

\begin{def.}
  Let $\langle \mathcal{S}, h \rangle$ be a state space topology with a set of states $S$.
  Let $s \in S$ be a progress state.

  The {\it bench level} of $s$ is $level(s) = \hwm(\Succ(s))$.

  The {\it inner bench states} $inner(s)$ for $s$ consist of all states $s'' \neq s$
  that can be reached from $s$ on paths
  on which all states $s' \neq s$ (including $s''$ itself) are non-progress states
  and satisfy $h(s') \leq level(s)$.

  The {\it bench exit states} $exit(s)$ for $s$ consist of all progress states $s'$ with $h(s') = level(s)$
  that are successors of $s$ or of some inner bench state of $s$.

  The {\it bench states} $states(s)$ for $s$ are $\{ s \} \cup inner(s) \cup exit(s)$.

  The {\it bench induced by} $s$, denoted by $\mathcal{B}(s)$, is the state space
  with states $states(s)$, initial state $s$, and goal states $exit(s)$.
  The successor function is the successor function of $S$ restricted to $states(s)$
  without transitions to $s$ and from bench exit states $exit(s)$.
  \label{def:bench}
\end{def.}

\begin{def.}
  Let $\mathcal{T} = \langle \mathcal{S}, h \rangle$ be a state space topology with initial state $s_{init}$.
  The bench transition system $\mathcal{B}(\mathcal{T})$ of $\mathcal{T}$ is a directed graph $\langle V, E \rangle$
  whose vertices are benches.
  The vertex set $V$ and directed edges $E$ are inductively defined
  as the smallest sets that satisfy the following properties:

  \begin{enumerate}
    \item $\mathcal{B}(s_{init}) \in V$
    \item if $\mathcal{B}(s) \in V$, $s' \in exit(s)$, and $s'$ is a non-goal state,
      then $\mathcal{B}(s') \in V$ and $\langle \mathcal{B}(s), \mathcal{B}(s') \rangle \in E$
  \end{enumerate}
  \label{def:bts}
\end{def.}

\begin{th.}\citep{HeusnerKH17}
  Let $\mathcal{T} = \langle \mathcal{S}, h \rangle$ be a state space topology
  with set of states $S$ and bench transition system $\langle V, E \rangle$.
  For each state $s \in S$, it holds that $s \in \mathcal{B}(s')$ for some $\mathcal{B}(s') \in V$
  iff there is a tie-breaking strategy with which GBFS expands $s$. %
  \label{th:bts}
\end{th.}

\subsubsection{BTS-Constrained Search}
By Theorem~\ref{th:bts}, the bench transition system ($\BTS$) defines the set of all states which are candidates for expansion by GBFS with some tie-breaking strategy.

Restricting the search to only expand states which are in the $\BTS$ is a natural constraint for parallel GBFS.

\begin{def.}
  A search algorithm is {\it $\BTS$-constrained} if it expands only states which are in the $\BTS$ \cite{ShimodaF23}.
  \end{def.}

\subsubsection{PUHF: A BTS-Constrained Parallel GBFS}
Kuroiwa and Fukunaga (\citeyear{KuroiwaF2020}) proposed Parallel Under High-water mark First (PUHF), a $\BTS$-constrained parallel GBFS.
PUHF marks states which are guaranteed to be in the BTS as {\it certain}, and only expands states marked as certain.
The criterion used by PUHF to mark states as certain was a restrictive, sufficient (but not necessary) condition for being in the BTS.
Recently, looser sufficient conditions for marking states were proposed, resulting in PUHF2--4, which significantly improved performance over PUHF \cite{ShimodaF23}.

\subsubsection{Constrained vs. Unconstrained Parallel GBFS}
More generally,  we say that a best-first search algorithm is  {\it constrained} if it expands the best state $s$ in $\textit{OPEN}$ only if $s$ satisfies some additional constraint $C$.  For example, PUHF is constrained, as it expands the best state in $\textit{OPEN}$ only if it is in the BTS.
We say that a best-first search algorithm is  {\it unconstrained} if it unconditionally expands the best state $s$ in $\textit{OPEN}$.
 For example, KPGBFS is unconstrained.

 \subsubsection{SGE: Separate Generation and Evaluation}
 Unconstrained parallel algorithms such as KPGBFS will unconditionally expand the top states in $\open$, so threads will be kept fully busy as long as $\open$ is not empty.
In contrast, constrained algorithms such as PUHF
can only expand states when it is guaranteed that they satisfy some state expansion constraint.
Even if $\open$ is non-empty, threads will remain idle until a state which satisfies the expansion constraint becomes available.
This results in lower evaluation rates compared to unconstrained algorithms \cite{KuroiwaF2020,ShimodaF23}.

Separate Generation and Evaluation (SGE) \cite{ShimodaF24} is an implementation technique for parallel search which parallelizes state evaluations in addition to expansions. This allows better utilization of  threads which would otherwise be idle (waiting for a state which satisfies expansion constraints).
The main idea is to decompose the expansion of state $s$ into separate units of work which can be parallelized: 
(1) successor generation, which generates the $succ(s)$, the successors of $s$, and
(2) successor evaluation, which evaluates $succ(s)$.

    In SGE, after a thread selects a state for expansion from the shared $\open$, it generates $succ(s)$, and inserts $succ(s)$  into a shared $\Unevaluated$ queue.
    The evaluation of states in $\Unevaluated$ is done in parallel, taking precedence over selection of states for expansion (a thread will select a state for expansion from $\open$ only if $\Unevaluated$ is currently empty). %
    Evaluated states are {\it not} immediately inserted into $\open$. Instead, SGE inserts all members of $succ(s)$ of $s$ simultaneously into $\open$, after they have all been evaluated %
    This is so that the parallel search is able to prioritize $succ(s)$ similarly to GBFS (otherwise, $succ(s)$ can be expanded in a completely different order than by GBFS).

\section{Pathological Behavior of PUHF}
\label{sec:puhf-pathology}
In this section, ``PUHF'' refers to PUHF \cite{KuroiwaF2020} and PUHF2-4 \cite{ShimodaF23} -- although these  PUHF variants use slightly different state expansion constraints, the arguments and conclusions below are unaffected by these differences.

While PUHF is guaranteed to only expand states in the BTS, previous work did not guarantee a bound on the number of states expanded relative to sequential GBFS.
We show that in fact PUHF is pathological relative to sequential GBFS.
The example we use in the proof motivates OBAT, the new bounded approach we propose in Section \ref{sec:bepuhf}.

\par
\begin{figure}[tb]
  \centering
  \includegraphics[width=0.98\columnwidth,,clip]{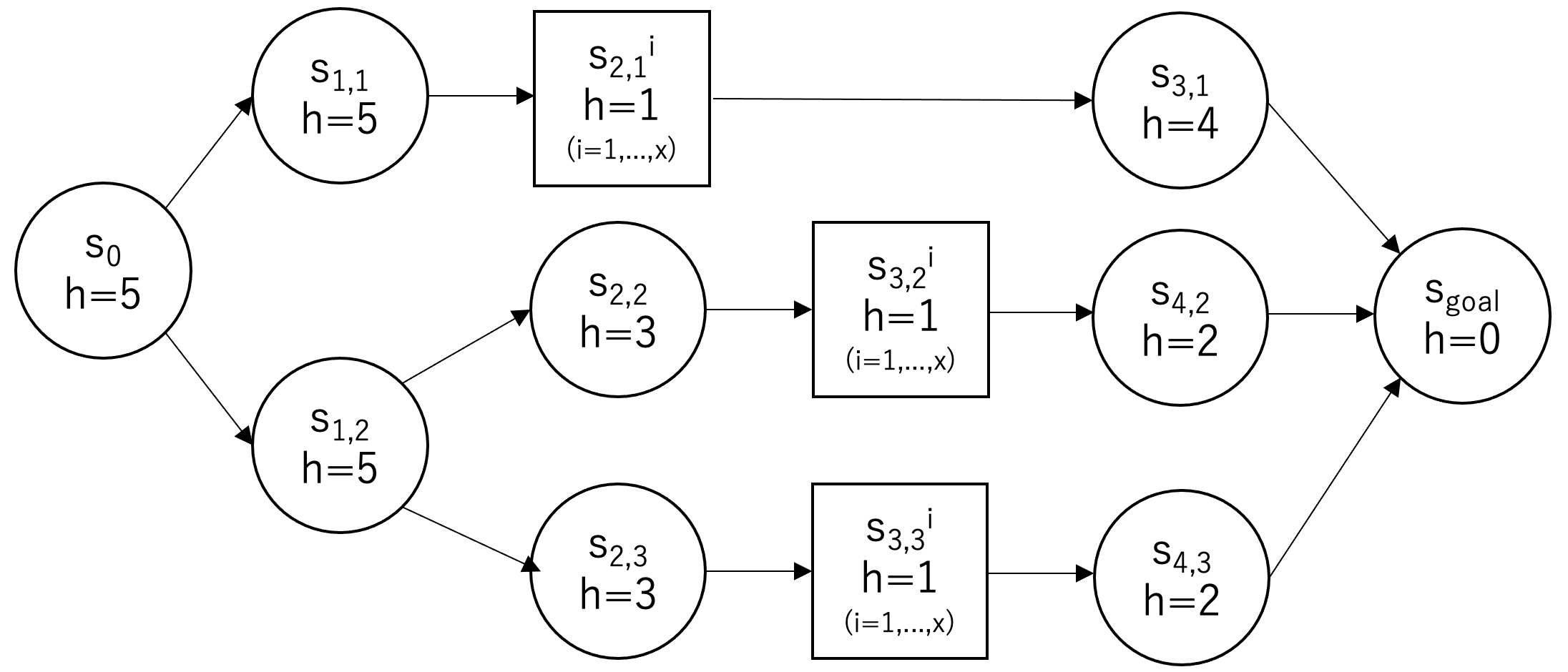}
  \caption{PUHF Pathological Search Behavior Example 1}
  \label{fig:t-pathological}
\end{figure}
\begin{figure}[tb]
  \centering
  \includegraphics[width=0.98\columnwidth,,clip]{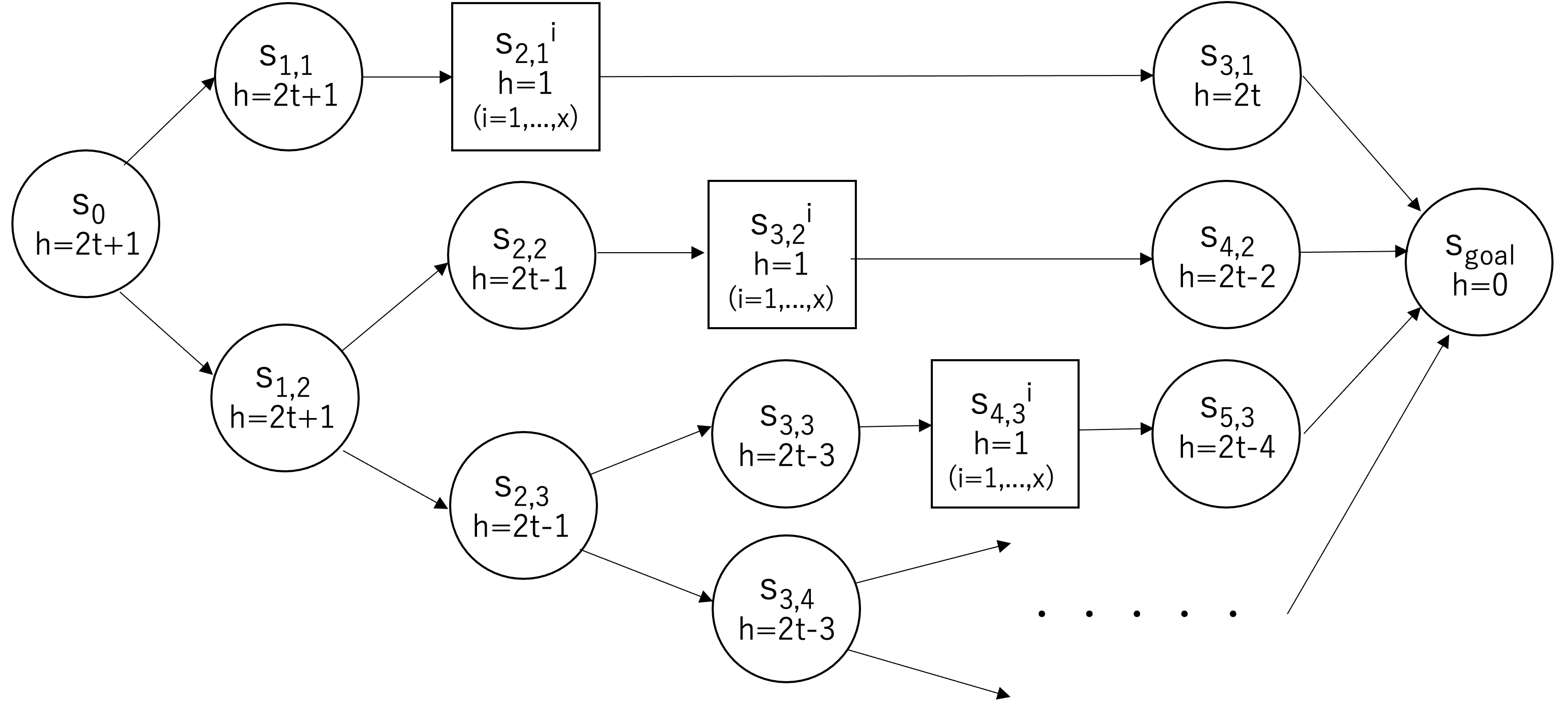}
  \caption{PUHF Pathological Search Behavior Example 2}
  \label{fig:t-pathological2}
\end{figure}

  Consider the search space in Figure~\ref{fig:t-pathological}.
  GBFS, under any tie-breaking policy must expand either
  $s_0 \rightarrow s_{1,1} \rightarrow {\bf s_{2,1}^{i}} \rightarrow s_{3,1} \rightarrow s_{goal}$, $s_0\rightarrow s_{1,2} \rightarrow s_{2,2}\rightarrow {\bf s_{3,2}^{i}} \rightarrow s_{4,2} \rightarrow s_{goal}$, 
  or 
  $s_0\rightarrow s_{1,2} \rightarrow s_{2,3}\rightarrow {\bf s_{3,3}^{i}} \rightarrow s_{4,3} \rightarrow s_{goal}$ 
  where ${\bf s^i}$ denotes a line of $x$ states.
  \par
  However, PUHF with $k \geq 2$ ($k$ is the number of threads) may (depending on tie-breaking) expand states in the order: 
  $s_0 \rightarrow (s_{1,1} \& s_{1,2}) \rightarrow {\bf s_{2,1}^{i}} \rightarrow (s_{2,2} \& s_{2,3})\rightarrow ({\bf s_{3,2}^{i}} \cup {\bf s_{3,3}^{i}}) \rightarrow (s_{4,2} \& s_{4,3})\rightarrow s_{goal} $
  where $a \& b$ denotes that states $a$ and $b$ are simultaneously expanded.
  All of these states are in the BTS, and when they are expanded in this order, they satisfy the expansion criteria used by all previous variants of PUHF.
  
  PUHF expands $3x$ states (denoted by the boxes $s_{2,1}$, $s_{3,2}$, $s_{3,3}$), as well as the 6 other non-goal states. For large $x$, PUHF with $k\geq 2$ may expand 3 times the number of states expanded by GBFS with the worst-case tie-breaking policy.
  \par
  By recursively nesting the structure of the search space in Figure~\ref{fig:t-pathological},
  we can construct the search space of Figure~\ref{fig:t-pathological2}.
  On this search space, following the reasoning above for Figure~\ref{fig:t-pathological},
  GBFS with the worst-case tie-breaking policy will expand $t+x+c$ states ($c$ is a constant).
  In contrast, PUHF may expand more than $x \cdot t$, where $t$ can be made arbitrarily large.
  Thus, PUHF with $k\geq 2$ and any tie-breaking strategy is pathological relative to GBFS.
  PUHF is pathological relative to GBFS even if the heuristic is consistent.
  For example, consider an instance of the search space with the structure shown in Figure~\ref{fig:t-pathological2}, where all edge costs are 1. If we replace all $h$-values by $h'$, such that $h' = h/c$, where $c$ is a sufficiently large constant such that $h/c < 1$, then  $h'$ is consistent, but PUHF is pathological on this search space.

Why PUHF is pathological relative to GBFS even though it is BTS-constrained is that unlike GBFS, parallel algorithms such as PUHF can explore multiple benches simultaneously.
\par
In Figure~\ref{fig:t-pathological}, $s_{1,1}$ and $s_{1,2}$ are both progress states. The benches entered through $s_{1,1}$ and $s_{1,2}$ are both part of the BTS, and are candidates for exploration by GBFS.
GBFS will expand either $s_{1,1}$ or $s_{1,2}$ and explore one of the corresponding benches,
and cannot explore the other bench until exiting the first bench.
On the other hand, PUHF can expand both $s_{1,1}$ and $s_{1,2}$, causing both benches to be explored.
As a result, although PUHF only expands states in the BTS, it can explore sets of benches which are not explored by GBFS with any one specific tie-breaking strategy.

\section{One Bench At a Time (OBAT)}
\label{sec:bepuhf}

\par
We now propose One Bench At a Time (OBAT), a BTS-constrained parallel BFS which behaves similarly to GBFS in that it is constrained to explore one bench at a time.
The main idea is to enforce this constraint by carefully controlling the expansion of potential progress states so that if multiple potential progress states are expanded simultaneously, only the successors of one of these potential progress states are allowed to be inserted into $\open$.
For example, if $s_1$ and $s_2$ are expanded simultaneously,
and they are both potential progress states, then only the successors of $s_1$ are inserted into $\open$, because if both $succ(s_1)$ and $succ(s_2)$ are inserted into $\open$, we may start to explore multiple benches simultaneously.

OBAT classifies states into two categories, $a$-states and $b$-states.
A state $s$ is an $a$-state if it does not have any successor with a lower $h$-value than $h(s)$. An $a$-state cannot be a progress state
(because if $s' \in succ(s)$ and $h(s) \leq  h(s')$, then $\hwm(s) \leq hwm(s')$, so if no successor of $s$ has a lower $h$-value than $s$, then $s$ cannot satisfy Definition \ref{def:progress}).
A state $s$ is a $b$-state if it has a successor with a lower $h$-value than $h(s)$. A $b$-state is potentially a progress state.
It is possible to determine whether a state is an $a$-state or $b$-state immediately after its successors are generated and their $h$-values are evaluated.

If $s$ is an  $a$-state, OBAT proceeds normally as with standard GBFS, inserting its successors in $\open$.
However, if $s$ is a $b$-state, it is a potential progress state, so instead of inserting the successors of $s$ into $\open$, it inserts $s$ into $\Deferred$, a priority queue (priority function is $h$-value).
A state $s$ in $\Deferred$ is ``on hold'' -- its successors are not inserted in $\open$ until $s$ is removed from $\Deferred$.
Thus, if multiple $b$-states are simultaneously expanded from $\open$,
$\Deferred$ prevents their successors from being simultaneously inserted into $\open$.

The sequence in which a state moves among the data structures in OBAT (i.e., its ``life cycle'') depends on whether it is an $a$-state or $b$-state.
$a$-states are: (a1) generated and evaluated, (a2) inserted in $\open$ and $\closed$, (a3) removed from $\open$.
$b$-states are: (b1) generated and evaluated, (b2) inserted in $\open$ and $\closed$, (b3) removed from $\open$, (b4) inserted in $\Deferred$, (b5) removed from $\Deferred$.

\begin{def.}
  A  state is {\it committed} after it leaves $\open$ or $\Deferred$ for the last time. Specifically, $a$-states are committed after (a3), and $b$-states are committed after (b5).
 \end{def.}

\begin{def.}
  A state $s$ is {\it completely expanded} if it is committed and  $succ(s)$ have all been inserted in $\open$.
\end{def.}

\begin{algorithm}[tb]
    \begin{algorithmic}[1]
        \scriptsize
        \State {$\open \leftarrow \{s_{init}\}, \closed\leftarrow\{s_{init}\}; {\forall{i}, s_i \leftarrow{\NULL}}$}
        \For{$i \leftarrow {0, ..., k-1}$ in parallel}
        \Loop
            \While{$s_i=\NULL$}
            \State {lock($\open$), lock($\Deferred$)}
            \If{$\open = \emptyset \, and\, \Deferred= \emptyset$}
                \If {$\forall{j}, s_j=\NULL$} 
                \State {unlock($\open$), unlock($\Deferred$); \Return $\NULL$}
            \EndIf
            \Else
                \If {$h(top(\Deferred)) \leq h(top(\open))$} \label{alg:OBAT:Deferred_pop}
                    \If {$h(top(\Deferred))\leq{min_{0\leq{j}<k}h(s_j)}$} \label{alg:OBAT:PUHF2criterion1}
                        \State $s \leftarrow top(\Deferred)$ \label{alg:OBAT:Open_t_pop}
                        \State $\Deferred \leftarrow \Deferred \backslash \{s\}$ 
                        \State lock($\closed$)
                        \For{$s' \in succ(s)$} \label{alg:OBAT:Deferred_insert}
                        \If{$s' \notin \closed$}
                        \State $\closed \leftarrow{\closed \cup \{s'\}}$
                        \State $\open \leftarrow{\open \cup \{s'\}}$
                        \EndIf
                        \EndFor       
                        \State unlock($\closed$) \Comment{$s$ is completely expanded}           
                    \EndIf
                \Else \label{alg:OBAT:Open_pop}
                    \If {$h(top(\open))\leq{min_{0\leq{j}<k}h(s_j)}$} \label{alg:OBAT:PUHF2criterion2}
                    \State $s_i \leftarrow top(\open);\open \leftarrow \open \backslash \{s_i\}$ \label{alg:OBAT:pop}
                    \EndIf
                \EndIf
            \EndIf
            \State {unlock($\open$), unolock($\Deferred$)}
            \EndWhile

            \If {$s_i \in S_{goal}$}
              \State \textbf{return} $Path(s_i)$
            \EndIf
            \If{$h(s_i)\leq min(h(succ(s_i)))$} \Comment{$s_i$ is an $a$-state}\label{alg:OBAT:normal expansion}
                \State lock($\open$), lock($\closed$)
                \For{$s_i' \in succ(s_i)$}
                  \If{$s_i' \notin \closed$}
                    \State $\closed \leftarrow{\closed \cup \{s_i'\}}$
                    \State $\open \leftarrow{\open \cup \{s_i'\}}$
                  \EndIf
                  \EndFor
                \State unlock($\open$), unlock($\closed$)
                \Comment{$s_i$ is completely expanded}      
            \Else \Comment{$s_i$ is a $b$-state}\label{alg:OBAT:virtual expansion}
              \State lock($\Deferred$)
              \State $\Deferred \leftarrow{\Deferred \cup \{s_i\}}$
              \State {unlock($\Deferred$)}
            \EndIf
            \State {$s_i \leftarrow{\NULL}$}
        \EndLoop
        \EndFor
        \caption{One Bench At a Time (OBAT)}
        \label{alg:OBAT}
    \end{algorithmic}
\end{algorithm}

\par
Algorithm \ref{alg:OBAT} shows the pseudocode for OBAT.
$top(Q)$ returns a highest priority (minimal $h$-value, ties broken according to a tie-breaking policy) state from queue $Q$. 
By convention, $h(top(Q)) = \infty$ if $Q$ is empty.
For all threads in parallel, OBAT proceeds as follows.

When $h(top(\Deferred)) \leq  h(top(\open))$ (line~\ref{alg:OBAT:Deferred_pop}) and $h(top(\Deferred))$ is less than or equal to the minimal $h$-value among all states currently being expanded by other threads (line~\ref{alg:OBAT:PUHF2criterion1}),
we pop $top(\Deferred)$ and insert its successors into $\open$, completing its expansion.

If $h(top(\open)) < h(top(\Deferred))$ (line~\ref{alg:OBAT:Open_pop}) and
$h(top(\open))$ is less than or equal to the smallest $h$-value among states currently being expanded by other threads (line~\ref{alg:OBAT:PUHF2criterion2}),
we set $s_i$ to $top(\open)$, pop $top(\open)$ (line~\ref{alg:OBAT:pop}), and generate $succ(s_i)$.

If $s_i$ is an $a$-state (line~\ref{alg:OBAT:normal expansion}), $succ(s_i)$ are inserted into $\open$ (so $s_i$ is completely expanded).
If $s_i$ is a $b$-state (line~\ref{alg:OBAT:virtual expansion}), $s_i$ is inserted into $\Deferred$, and $succ(s_i)$ are {\it not} inserted anywhere, but they remain linked to $s_i$.

Lines \ref{alg:OBAT:PUHF2criterion1} and \ref{alg:OBAT:PUHF2criterion2} ensure that OBAT will not expand states with a higher $h$-value than any state which is currently expanded by another thread. This is the same expansion constraint as PUHF2, so it follows directly from \citep[Theorem 4]{ShimodaF23} that:
\begin{prop.}
  OBAT is BTS-constrained.
\end{prop.}

\subsubsection{Example of OBAT Search Behavior}

To illustrate the behavior of OBAT, consider an example run of OBAT  with 2 threads on the search space in Figure~\ref{fig:t-pathological}.
First, one thread expands the initial state $s_0$. Since $s_0$ does not have any successors with a smaller $h$-value,
$s_{1,1}$ and $s_{1,2}$ are both inserted in $\open$.
Next, threads 1 and 2 remove $s_{1,1}$ and $s_{1,2}$ from $\open$, respectively and expand them.
Since $s_{1,1}$ and $s_{1,2}$ both have successors with a smaller $h$-value than themselves, $s_{1,1}$ and $s_{1,2}$ are both potential progress states, so they are both inserted into $\Deferred$.
Next, a state is removed from $\Deferred$. Assume $s_{1,1}$ is removed (the specific state depends on the tie-breaking policy).
$s_{1,1}$ is completely expanded ($succ(s_{1,1})$ are inserted into $\open$).
Then, both threads will search the line of states  $s_{2,1}^1,..,s_{2,1}^x$, and then $s_{3,1}$, reaching the goal.
Thus, whenever OBAT selects for expansion a potential progress state ($s_{1,1}$ and $s_{1,2}$ in this example),
OBAT defers them, and will only completely expand one of them,
thereby preventing simultaneous exploration of multiple benches.

\section{Analysis of State Expansions by OBAT}
\label{sec:bepuhf-analysis}
We now compute a bound on the number of states expanded by OBAT.
The number of states expanded by OBAT is the sum of (1) the number of completely expanded states and (2)  the number of states remaining in $\Deferred$ when the algorithm terminates.

\begin{th.}
  The number of states completely expanded by OBAT is less than or equal to the number of states expanded by GBFS with some tie-breaking policy.
\label{th:OBAT_OPEN}
\end{th.}
\begin{proof}
  Let $s_1$,$s_2$,$s_3$... be a serialized ordering of the states which are completely expanded by OBAT, in the order in which they are committed. %
  We show that there exists a tie-breaking policy for sequential GBFS which expands states in this same order. The proof is by induction.
  For both sequential GBFS and OBAT, the first (completely) expanded state $s_1$ corresponds to the initial state $s_{init}$.
  Next, assume that for the ordering $s_1,..., s_i$, there exists a tie-breaking policy for GBFS which expands states in that order.
  We show that there exists a tie-breaking policy for sequential GBFS which will expand $s_{i+1}$ after $s_i$.

  When sequential GBFS removes $s_{i+1}$ from $\open$,  the set of states in $\open$ are
  $\bigcup_{k=1}^{k=i}{succ(s_k)} \setminus \bigcup_{k=1}^{k=i}{s_k}$ (the set of all generated states minus the set of all expanded states).
  Now consider when OBAT commits $s_{i+1}$. %
  Let $A$ and $B$ denote the set of $a$-states and $b$-states currently being expanded by other threads. Let $A'$ denote the set of successors of all states in $A$.
  $\bigcup_{k=1}^{k=i}{succ(s_k)} \setminus \bigcup_{k=1}^{k=i}{s_k} = (\open \cup \Deferred \cup B \cup  (A' \setminus \bigcup_{k=1}^{k=i}{s_k}))$.
  $s_{i+1}$ has the smallest $h$-value among all states in $\Deferred \cup \open$.
  
  $A \cup B$ cannot contain any states taken from $\Deferred$, because $\open$ and $\Deferred$ are locked while a thread removes a state from $\Deferred$ and inserts its successors into $\open$, so this cannot occur simultaneously with the complete expansion of $s_{i+1}$, which requires access to $\open$ and $\Deferred$.
  Thus, all states in $A \cup B$ were taken from $\open$.
  All states in $B$ will later be inserted in $\Deferred$, so are not in $s_1,...s_i$.
  All states in $A$ will be included in $s_1,...,s_i$. By definition, $a$-states do not have any successors with a smaller $h$-value than itself.
  
  Due to lines  \ref{alg:OBAT:PUHF2criterion1} and \ref{alg:OBAT:PUHF2criterion2},
  $h(s_{i+1})\leq h(c) \forall c \in (A \cup B)$.
  Therefore, $h(s_{i+1}) \leq min_{a \in A} h(a)  \leq min_{u \in A'} h(u)$.
  Thus, $h(s_{i+1})$ has the smallest $h$-value among $(\open \cup \Deferred \cup B \cup ( A' \setminus \bigcup_{k=1}^{k=i}{s_k}))$.

  This implies, $s_{i+1}$ has the smallest $h$-value in $\open$ after GBFS expands $s_i$, so there exists a tie-breaking policy for sequential GBFS which will select $s_{i+1}$ for expansion immediately after $s_{i}$.
  Thus, there exists a tie-breaking policy for GBFS which expands states in the same order as they are committed in OBAT. %
  Therefore, the number of states completely expanded by OBAT is less than or equal to the number of states expanded by GBFS with the worst-case tie-breaking policy.
\end{proof}

To bound the number of states in $\Deferred$ when OBAT terminates,
we bound the number of different $h$-values for which there is a state in $\Deferred$ (Lemma~\ref{lem:Open_t_kind}) and the number of states per $h$-value (Lemma~\ref{lem:Open_t_num}).

\begin{lem.}
  When OBAT terminates after finding a solution path $p$, the number of $h$-values for which at least one state is in $\Deferred$ is at most $|p|$.
  \label{lem:Open_t_kind}
\end{lem.}

  \begin{proof}
  Let $p = {s_1,s_2,...,s_n}$ be the sequence of all states on the solution path (in the order they appear in the solution path), such that all states after $s_i$ on the solution path have a smaller $h$-value than $s_i$.
  From the definition of $p$,  all states in the solution path between $s_i$ and $s_{i+1}$ have an $h$-value less than or equal to $h(s_{i+1})$.
  Thus, after the successors of $s_i$ are inserted in $\open$, there is a state with $h$-value less than or equal to $h(s_{i+1})$ in $\open$, $\Deferred$, or a state currently being expanded by some thread.
  OBAT will not select a state from $\open$ with a $h$-value larger than a state currently being expanded by some thread   (lines \ref{alg:OBAT:PUHF2criterion1} and \ref{alg:OBAT:PUHF2criterion2}),
  so between the time after the successors of $s_i$ are inserted in $\open$ and $s_{i+1}$ is completely expanded, no states with $h$-values larger than $h(s_{i+1})$ will be removed from $\open$.

  States with $h$-value less than $h(s_{i+1})$ will have higher priority for expansion than $s_{i+1}$, so no such state can be in $\Deferred$ when $s_{i+1}$ is expanded.
  Thus, in the time between when $s_i$ is completely expanded, and when $s_{i+1}$ is removed from $\Deferred$ and committed, the only states which can be 
  removed from $\open$ and inserted into $\Deferred$ (and can remain in $\Deferred$ after OBAT terminates) have an $h$-value of $h(s_{i+1})$.
  Similarly, for the base case ($s_1$), the only states which can be removed from $\open$ before $s_1$ and might remain in $\Deferred$ after OBAT terminates have a $h$-value of $h(s_1)$.
  Therefore, the only possible $h$-values of states in $\Deferred$ when OBAT terminates are $h(s_1), h(s_2), ..., h(s_n)$, which is  $\leq n$ unique values.
\end{proof}
\begin{lem.}
  At all times, for every $h$-value, $\Deferred$ contains at most $k$ states, where $k$ is the number of threads.
  \label{lem:Open_t_num}
\end{lem.}

\begin{proof}
  Assume that $\Deferred$ contains $k+1$ states which all have the same $h$-value.
  By the pigeonhole principle, there must be 2 such states $s_1$ and $s_2$,
  which were inserted into $\Deferred$ by the same thread $T$, where $s_1$ was inserted before $s_2$.
  States in $\Deferred$ come from $\open$, i.e., a state is inserted in $\Deferred$ only after it is removed from $\open$.
  Therefore, for $s_2$ to be inserted into $\Deferred$, it must first be removed from  $\open$,
  and by assumption, at that time, $s_1$ is already in $\Deferred$.
  However, if $h(top(\open)) = h(top(\Deferred))$, then OBAT will remove a state from $\Deferred$ (Algorithm \ref{alg:OBAT}, line~\ref{alg:OBAT:Open_t_pop}), so $s_2 \in \open$ will not be removed before $s_1 \in \Deferred$, a contradiction.
\end{proof}

Lemma~\ref{lem:Open_t_kind} and Lemma~\ref{lem:Open_t_num} directly give us a bound on the number of states in $\Deferred$ when OBAT terminates.

\begin{th.}
  When OBAT terminates after finding a solution path $p$, the number of states which remain in $\Deferred$ is less than or equal to  $k|p|$, where $k$ is the number of threads.
  \label{th:OBAT_Deferred}
\end{th.}

Finally, from Theorem~\ref{th:OBAT_OPEN} and Theorem~\ref{th:OBAT_Deferred}, we have the following bound on the number of states expanded: 
\begin{th.}
  The number of states expanded by OBAT is at most $N_{\textit{gbfs}}$ + $k|p|$, where $N_{\textit{gbfs}}$ is the number of states expanded
  and $p$ is the solution path found by GBFS
  with the tie-break policy which maximizes $N_{\textit{gbfs}}$ + $k|p|$.
  \label{th:OBAT_total}
\end{th.}  

As an aside, since $|p| \leq N_{\textit{gbfs}}$, %
OBAT is $t$-bounded relative to GBFS   with the tie-break policy which maximizes $N_{\textit{gbfs}}$ + $k|p|$,  where $t=k+1$. 
However, $N_{\textit{gbfs}}$ is usually much larger than  $|p|$, so the bound in Theorem~\ref{th:OBAT_total} is more meaningful than the $t$-boundedness.

\section{\OBATS: OBAT with SGE}
\label{sec:obats}

Like PUHF and its variants, OBAT is a constrained parallel GBFS,
and threads can be forced to be idle while they wait until a
state which is guaranteed to satisfy the expansion constraint becomes available, resulting
in worse performance than unconstrained algorithms such as KPGBFS which never leave threads idle.
As with PUHF, this issue can be alleviated using Separate Generation and Evaluation \cite{ShimodaF24}, an implementation technique which decouples successor generation and evaluation to reduce idle waiting.

Applying SGE to OBAT is straightforward.
Algorithm \ref{supp:alg:OBATS} in the Supplement shows \OBATS, which is OBAT with SGE.
As SGE does not change the set of states which are possibly expanded, it can be proven straightforwardly that the  upper bound on the number of states expanded by OBAT (Theorem~\ref{th:OBAT_total}) also holds in \OBATS.

\section{Experimental Evaluation}
\label{sec:bepuhf-experiment}

We evaluated the performance of OBAT and \OBATS with 4, 8, 16 threads.
As baselines, we also evaluate KPGBFS, \KPDGBFS (KPGBFS with SGE), PUHF3, \PDUHF (PUHF3 with SGE) , and single-thread GBFS.

\subsubsection{Experimental Settings}
We compared the algorithms using a set of instances based on the Autoscale-21.11 satisficing benchmark set (42 STRIPS domains, 30 instances/domain, 1260 total instances) \cite{torralba-et-al-icaps2021}, an improved benchmark suite based on the IPC classical planning benchmarks.
However, for domains where  (1) all methods solved all instances (i.e., the instances were too easy for the purpose of comparing the performance of KPGBFS vs. OBAT), and (2) an instance generator for the domain is available in the Autoscale repository, we replaced the Autoscale-21.11 instances with more difficult instances generated using the Autoscale generator.
Specifically, the domains where criteria (1) and (2) above applied were 
\pddl{gripper} and \pddl{miconic}.
See Supplement \ref{supp:benchmark selection} for additional details, including instance generator parameters.

All algorithms use the FF heuristic \cite{Hoffmann01}. %
Each run had a time limit of 5 min.,  3GB RAM/thread (e.g., 8 threads: 24GB total) limit on an Intel Xeon CPU E5-2670 v3@2.30GHz processor.
All tie-breaking is FIFO.
The code is available at \url{https://github.com/TakuShimoda/AAAI25}.

\begin{table}[tbh]%
  \centering
    {\scriptsize
  \begin{tabular}{|l|r|r|r|r|}
    \hline
    \#threads &  1 thread & 4 threads & 8 threads & 16 threads \\
    \hline
    GBFS  & 401 & \multicolumn{3}{|c|}{-} \\    
    \hline
    KPGBFS  & - & 462 & 488 & 529\\
    \KPDGBFS  & - &472 & 500 & {\bf 532} \\
    PUHF3 & - &  459 &  477 & 494 \\
    \PDUHF & - &  468 & 494 & 510 \\
    \hline
    OBAT &  - & 458 & 477 & 496 \\
    \OBATS &  - & {\bf 478} & {\bf 506} & {\bf 532} \\
    \hline
  \end{tabular}
  }
  \caption{Coverage (number of problems solved out of 1260) }
  \label{tab:coverage}
\end{table}

\begin{table}[tbh]
  \centering
    {\scriptsize
  \begin{tabular}{|l|r|r|r|r|r|r|r|r|r|r|r|r|r|r|r|r|r|r|r|r|}
    \multicolumn{1}{c}{\rot {}} & \multicolumn{1}{c}{\rot {GBFS}}  & \multicolumn{1}{c}{\rot {KPGBFS}} & \multicolumn{1}{c}{\rot {KPGBFS$_S$}} & \multicolumn{1}{c}{\rot {PUHF3}} & \multicolumn{1}{c}{\rot {PUHF3$_S$}} &  \multicolumn{1}{c}{\rot {OBAT}} & \multicolumn{1}{c}{\rot {OBAT$_S$}} \\
    \hline
    agricola & 26 & 30 & 30 & 30 & 30 & 30 & 30 \\
    airport & 18  & 17 & 18 & 17 & \textbf{19} & 15 & 18 \\
    barman & 2 &  4 & 4 & 4 & 4 & 4 & \textbf{5} \\
    blocksworld & 6 & 7 & 7 & 7 & \textbf{8} & 6 & 7 \\
    childsnack & 4 & \textbf{5} & 3 & \textbf{5} & 4 & 4 & 4 \\
    data-network & 4 & \textbf{7} & 6 & 6 & 4 & 5 & 5 \\
    depots & 4 & \textbf{5} & \textbf{5} & 4 & 4 & 4 & 4 \\
    driverlog & 4 & 7 & 8 & 8 & 8 & 7 & \textbf{9} \\
    elevators & 11 & 18 & 19 & \textbf{20} & \textbf{20} & \textbf{20} & 19 \\
    floortile & 2 &2 & 2 & 2 & 2 & 2 & 2 \\
    \hline
    freecell & 20 & 24 & 23 & 21 & 22 & 22 & \textbf{25} \\
    ged & 7 & 9 & 9 & 7 & 8 & \textbf{10} & 8 \\
    grid & 5 & 5 & \textbf{6} & \textbf{6} & \textbf{6} & \textbf{6} & \textbf{6} \\
    gripper (replaced) & 2 & 7 & \textbf{21} & 7 & 20 & 19 & 20 \\
    hiking & 4 & \textbf{14} & 9 & 11 & 8 & 10 & 9 \\
    logistics & 4 & \textbf{6} & \textbf{6} & 5 & 5 & \textbf{6} & \textbf{6} \\
    miconic (replaced) & 10 & 9 & 15 & 9 & 15 & 10 & \textbf{21} \\
    mprime & 4 & \textbf{6} & \textbf{6} & \textbf{6} & \textbf{6} & 5 & 5 \\
    nomystery & 6 & \textbf{14} & 11 & 11 & 11 & 3 & 6 \\
    openstacks & 7 & 12 & 11 & 11 & 11 & 15 & \textbf{22} \\
    \hline
    organic-synthesis-split & 9 & \textbf{17} & \textbf{17} & 16 & 16 & 13 & 16 \\
    parcprinter & 30 & 30 & 30 & 30 & 30 & 30 & 30 \\
    parking & 6 & 9 & \textbf{10} & 9 & 9 & 9 & \textbf{10} \\
    pathways & 11 & \textbf{15} & 12 & 10 & 12 & 11 & 12 \\
    pegsol & 30 & 30 & 30 & 30 & 30 & 30 & 30 \\
    pipesworld-notankage & 8 & \textbf{11} & 9 & 9 & 9 & 9 & \textbf{11} \\
    pipesworld-tankage & 9 & \textbf{12} & \textbf{12} & 11 & 11 & \textbf{12} & \textbf{12} \\
    rovers & 26 & \textbf{27} & \textbf{27} & 26 & 26 & 26 & 26 \\
    satellite & 7 & 7 & 7 & 7 & 7 & 7 & 7 \\
    scanalyzer & 7 & 11 & 11 & 11 & 11 & 11 & 11 \\
    \hline
    snake & 7 & \textbf{10} & \textbf{10} & 9 & 9 & 9 & 9 \\
    sokoban & 16 & \textbf{23} & 22 & 21 & 21 & 19 & 19 \\
    storage & 3 & 2 & \textbf{4} & \textbf{4} & \textbf{4} & 3 & \textbf{4} \\
    termes & 12 & \textbf{18} & \textbf{18} & 16 & \textbf{18} & 15 & 16 \\
    tetris & 8 & \textbf{13} & 12 & \textbf{13} & 12 & \textbf{13} & \textbf{13} \\
    thoughtful & 14 & \textbf{21} & 18 & 19 & 16 & 16 & 14 \\
    tidybot & 12 & \textbf{14} & 13 & 12 & 11 & \textbf{14} & 13 \\
    tpp & 8 & 10 & 11 & 9 & 10 & 11 & \textbf{12} \\
    transport & 5 & \textbf{8} & 7 & \textbf{8} & 7 & 7 & 7 \\
    visitall & 14 & \textbf{20} & 19 & 13 & 14 & 15 & 16 \\
    woodworking & 2 & 2 & 2 & 2 & 2 & 2 & 2 \\
    zenotravel & 7 & 11 & \textbf{12} & \textbf{12} & 10 & 11 & 11 \\
    \hline
    Sum(1260) & 401 & 529 & \textbf{532} & 494 & 510 & 496 & \textbf{532} \\
    \hline
  \end{tabular}
  }
  \caption{Coverage (number of problems solved out of 1260) on Autoscale-21.11 IPC-based planning benchmark set (\pddl{gripper}, and \pddl{miconic} are replaced with harder instances. See \ref{supp:benchmark selection}) for $k$=16 threads (except $k$=1 for GBFS.}
  
  \label{tab:full coverage}
\end{table}

\subsubsection{Results}

\begin{figure}[tbh]
  \begin{subfigure}[]{0.32\columnwidth}
    \includegraphics[width=\textwidth,trim={1cm 0.25cm 2cm 0.25cm},,clip]{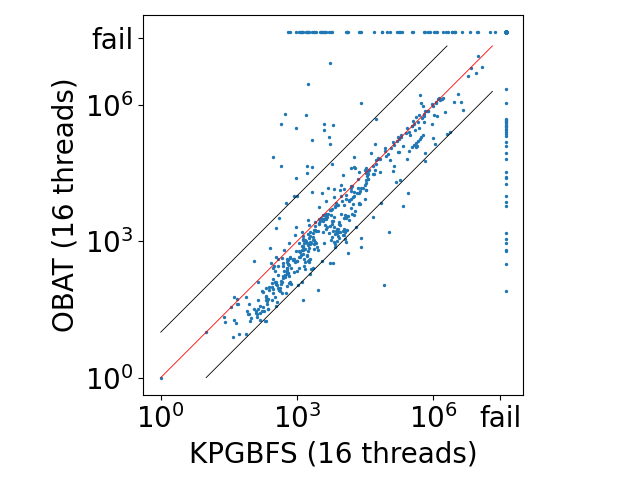}
  \end{subfigure}
  \begin{subfigure}[]{0.32\columnwidth}
    \includegraphics[width=\textwidth,trim={1cm 0.25cm 2cm 0.25cm},,clip]{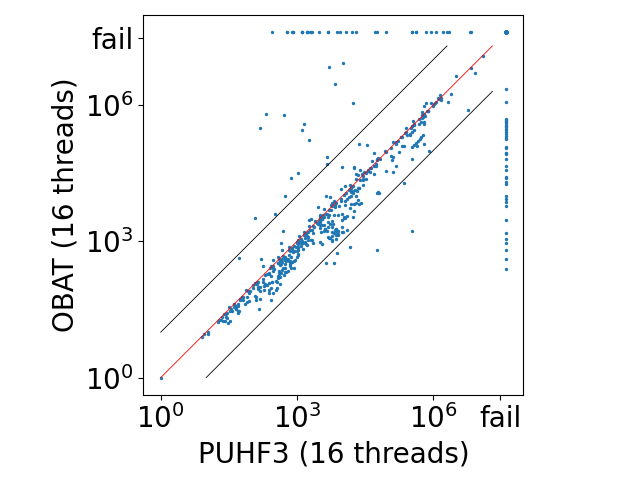}
  \end{subfigure}
  \centering
  \begin{subfigure}[]{0.32\columnwidth}
    \includegraphics[width=\textwidth,trim={1cm 0.25cm 2cm 0.25cm},,clip]{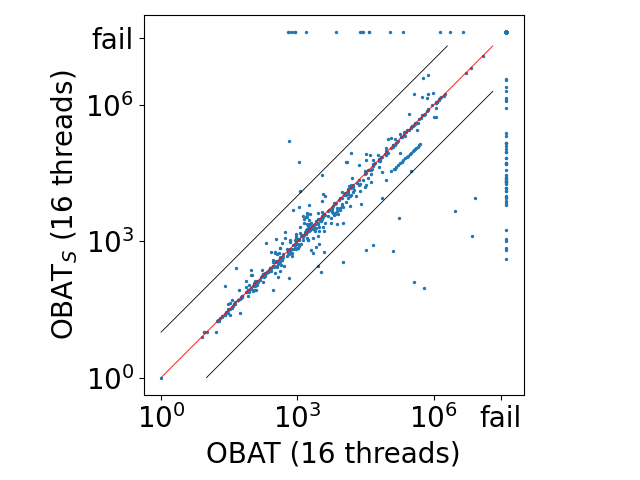}    %
  \end{subfigure}
  \centering
  \caption{Number of states expanded, 16 threads. Diagonal lines are $y=0.1x$, $y=x$, and $y=10x$}
  \label{fig:expansions-comparisons}
  
\end{figure}

\begin{figure}[tbh]
  \begin{subfigure}[]{0.32\columnwidth}
    \includegraphics[width=\textwidth,trim={1cm 0.25cm 2cm 0.25cm},,clip]{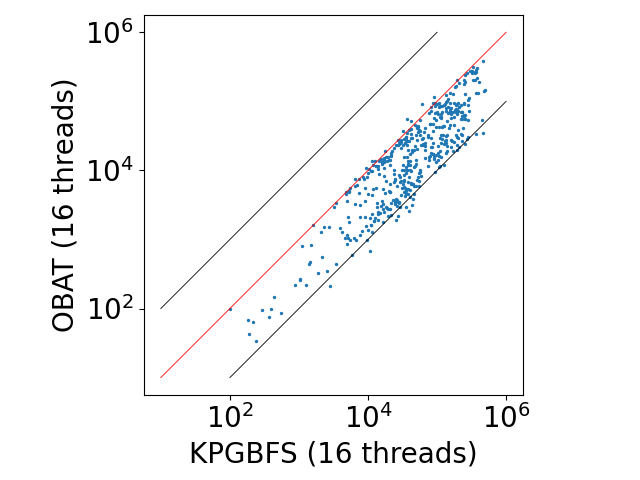}
  \end{subfigure}
  \begin{subfigure}[]{0.32\columnwidth}
    \includegraphics[width=\textwidth,trim={1cm 0.25cm 2cm 0.25cm},,clip]{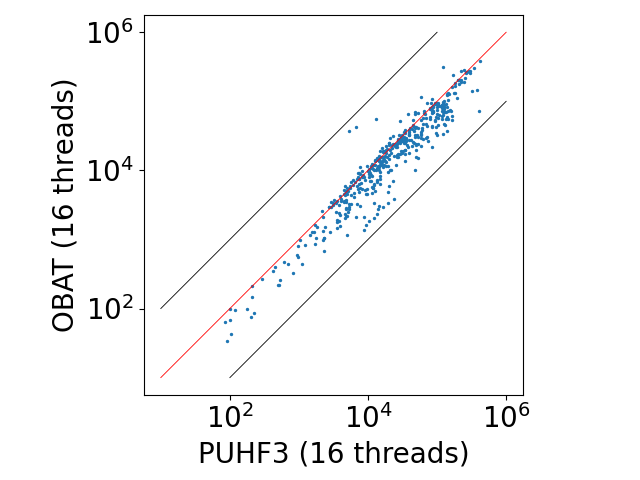}
  \end{subfigure}
  \centering

  \begin{subfigure}[]{0.32\columnwidth}
    \includegraphics[width=\textwidth,trim={1cm 0.25cm 2cm 0.25cm},,clip]{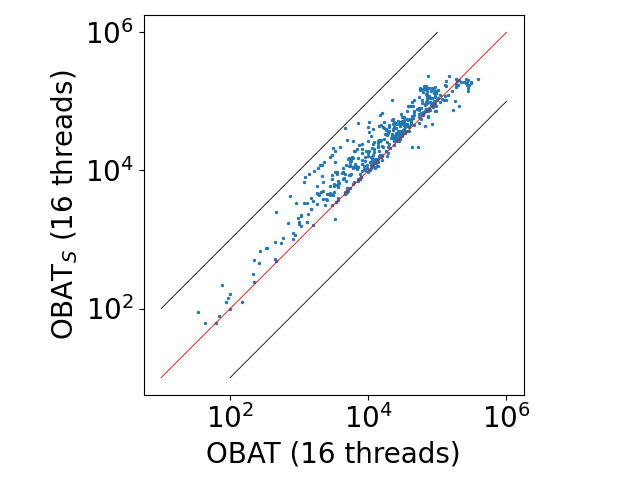}
  \end{subfigure}
  \begin{subfigure}[]{0.32\columnwidth}
    \includegraphics[width=\textwidth,trim={1cm 0.25cm 2cm 0.25cm},,clip]{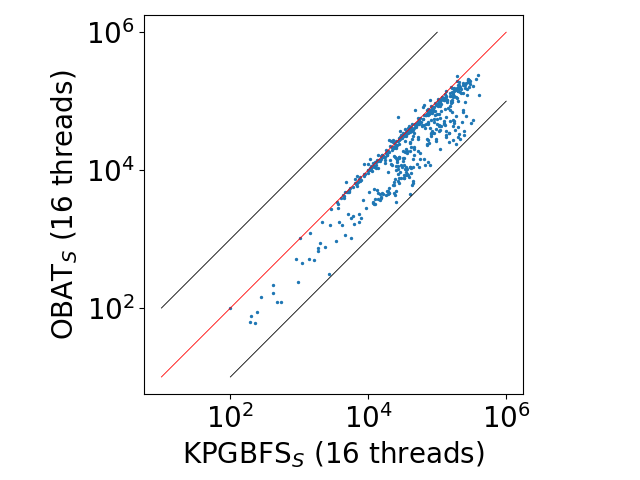}
  \end{subfigure}
  \begin{subfigure}[]{0.32\columnwidth}
    \includegraphics[width=\textwidth,trim={1cm 0.25cm 2cm 0.25cm},,clip]{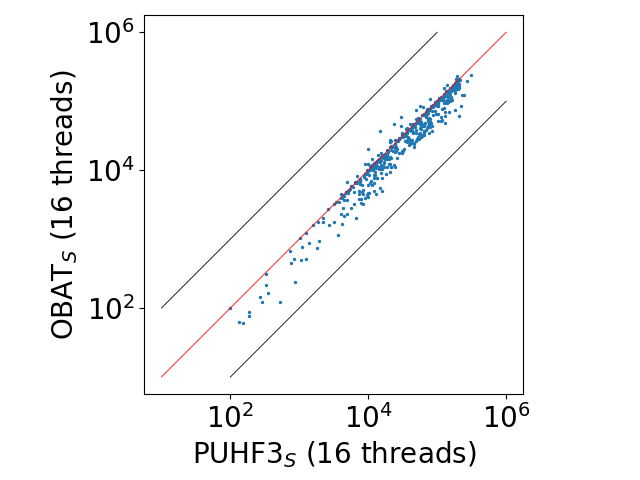}
  \end{subfigure}
  
  \centering
  \caption{ State evaluation rates (states/second), 16 threads. Diagonal lines are $y=0.1x$, $y=x$, and $y=10x$}
  \label{fig:evaluation_rate}

\end{figure}

Table~\ref{tab:coverage} shows the total coverage (number of instances solved out of 1260) for 4/8/16 threads.
Table~\ref{tab:full coverage} shows per-domain coverage for 16 threads.
Figure~\ref{fig:expansions-comparisons} compares the number of states expanded by the algorithms.
Figure~\ref{fig:evaluation_rate} compares the state evaluation rates of the algorithms.
Additional figures comparing number of state expansions, state evaluation rates, and search time including data for 4 and 8 threads, are in 
the Supplement, Figures~\ref{supp:fig:expansions-comparisons-OBAT}-\ref{supp:fig:search-time-comparisons-OBAT_S}.

Compared to KPGBFS, OBAT  expands significantly fewer states (Figure~\ref{fig:expansions-comparisons}, left).
However, OBAT has significantly lower state evaluation rates than KPBGFS (Figure~\ref{fig:evaluation_rate} top left). This results in significantly lower total coverage than KPGBFS (Tab.~\ref{tab:coverage}), although the difference is domain-dependent, e.g., OBAT significantly outperforms KPGBFS on \pddl{openstacks}.
Compared to PUHF3, OBAT also expands fewer states (Figure~\ref{fig:expansions-comparisons}, middle) has a lower evaluation rate (Figure~\ref{fig:evaluation_rate} top right), and has comparable total coverage.
Thus, the expansion constraint appears to enable OBAT to search good regions of the search space, at the cost of low evaluation rates.

\OBATS has a significantly higher evaluation rate than OBAT (Figure~\ref{fig:evaluation_rate}, bottom left) 
and a comparable number of state expansions (Figure~\ref{fig:expansions-comparisons}, right), showing that SGE significantly alleviates the drawback of OBAT without sacrificing search efficiency.
As a result, \OBATS achieves significantly higher coverage than OBAT (Table~\ref{tab:coverage}).

Overall, \OBATS achieves coverage competitive with all other methods for 4/8/16 threads,
showing that the guaranteed bound on the number of expansions does not require sacrifing practical performance.
As shown in Table~\ref{tab:full coverage}, none of the algorithms clearly dominate the others, and comparative performance is domain-dependent.
Despite the state evaluation rate boost due to SGE, \OBATS still has a significantly lower evaluation rate than \KPDGBFS and \PDUHF (Figure~\ref{fig:evaluation_rate}, bottom middle and right) -- further decreasing the evaluation rate gap (improving thread utilization)  while maintaining search efficiency is a direction for future work.

\section{Conclusion}
\label{sec:conclusion}
We proposed OBAT, a parallel GBFS algorithm which, to our knowledge, 
is the first parallel GBFS with a shared $\open$ list that guarantees a bound on the number of states expanded relative to sequential GBFS.
The worst-case gap in the number of state expansions between OBAT and sequential GBFS is the gap between the best-case sequential GBFS tie-breaking policy and the worst-case sequential GBFS tie-breaking policy.

These results establish a close relationship between parallel greedy best-first search and theoretical work in tie-breaking policies and opens interesting avenues for future work -- for example, any upper bound on the gap between best vs. worst-case tie-breaking will also imply an upper bound on the gap between sequential and parallel greedy best-first search.

We showed experimentally that while OBAT achieves search efficiency competitive with previous shared-$\open$ parallel GBFS algorithms, the state evaluation rate was significantly lower than previous algorithms.
Finally, we showed that applying SGE to OBAT significantly increases the state evaluation rate of OBAT, resulting in overall performance which is comparable to previous methods, at least up to 16 threads. Investigating and improving scalability to larger number of threads is a direction for future work.

\bibliography{references}
\par

\include{supplement-aaai25-arxiv}

\end{document}

%% file: supplement-aaai25-arxiv.tex
\onecolumn

\setcounter{section}{0}
\renewcommand{\thesection}{S.\arabic{section}}
\renewcommand{\thesubsection}{S.\arabic{section}.\arabic{subsection}}
\renewcommand{\thealgorithm}{S.\arabic{algorithm}}

\section{OBAT with SGE (OBAT$_S$)}

\begin{algorithm}[H]
    \begin{algorithmic}[1]
        \scriptsize
        \State {$\open \leftarrow \{s_{init}\}, \closed\leftarrow\{s_{init}\}; \forall{i}, s_i \leftarrow{NULL}$}
        \For{$i \leftarrow {0, ..., k-1}$ in parallel}
        \Loop
          \State lock($\Unevaluated$) \label{supp:alg:OBATS:Unevaluated start}
          \If{$\Unevaluated\neq\emptyset$}
            \State{$s_i \gets top(\Unevaluated)$}
            \State $Unevaluated \leftarrow \Unevaluated \backslash \{s_i\}$ 
            \State{unlock($\Unevaluated$)}
            \State{evaluate($s_i$)} \Comment{using a hashtable to prevent reevaluation of states}
            \State{$s \leftarrow parent(s_i)$}
            \If{all $succ(s)$ have been evaluated}
              \If{$h(s)\leq min_{s' \in succ(s)} h(s')$} \Comment{$s$ is an $a$-state}
                \State lock($\open$), lock($\closed$)      
                \For{$s' \in succ(s)$}
                  \If{$s' \notin \closed$}
                    \State $\closed \leftarrow{\closed \cup \{s'\}}$
                    \State $\open \leftarrow{\open \cup \{s'\}}$
                  \EndIf
                \EndFor
                \State unlock($\open$), unlock($\closed$) 
              \Else \Comment{$s$ is a $b$-state}
                \State lock($\Deferred$)
                \State $\Deferred \leftarrow{\Deferred \cup \{s\}}$
                \State unlock($\Deferred$)
              \EndIf
              \State{$Being\_Expanded \leftarrow Being\_Expanded \backslash \{s\}$}
            \EndIf  \label{supp:alg:OBATS:Unevaluated end}
          \Else
            \State {unlock($\Unevaluated$)} \label{supp:alg:OBATS:expansion start}
            \State {lock($\open$), lock($\Deferred$)}
            \If{$\open = \emptyset \, and\, \Deferred= \emptyset$}
              \State {unlock($\open$), unlock($\Deferred$)}
              \If {$Being\_Expanded = \emptyset$} \textbf{return} $NULL$ \EndIf
            \Else
              \If {$h(top(\Deferred)) \leq h(top(\open))$} \label{supp:alg:OBATS:expansion constraints1}
                \If {$h(top(\Deferred))\leq{min_{s \in Being\_Expanded}h(s)}$} \label{supp:alg:OBATS:expansion constraints2}
                  \State $s_i \leftarrow top(\Deferred); \Deferred \leftarrow \Deferred \backslash \{s_i\}$ \label{supp:alg: OBATS:expansion from Deffered start}
                  \State{$Being\_Expanded \leftarrow Being\_Expanded \cup \{s_i\}$}
                  \State lock($\closed$)
                  \For{$s_i' \in succ(s_i)$}
                    \If{$s_i' \notin \closed$}
                      \State $\closed \leftarrow{\closed \cup \{s_i'\}}$
                      \State $\open \leftarrow{\open \cup \{s_i'\}}$
                    \EndIf
                  \EndFor       
                  \State unlock($\closed$)
                  \State{$Being\_Expanded \leftarrow Being\_Expanded  \backslash \{s_i\}$}
                \EndIf \label{supp:alg: OBATS:expansion from Deffered end}
                \State unlock($\open$), unlock($\Deferred$)
              \Else 
                \If {$h(top(\open))\leq{min_{s \in Being\_Expanded}h(s)}$} \label{supp:alg:OBATS:expansion constraints3}
                  \State $s_i \leftarrow top(\open); \open \leftarrow \open \backslash \{s_i\}$ \label{supp:alg:OBATS:expansion from Open start}
                  \State{$Being\_Expanded \leftarrow Being\_Expanded \cup \{s_i\}$}
                  \State unlock($\open$), unlock($\Deferred$)
                  \If {$s_i \in S_{goal}$} \textbf{return} $Path(s_i)$ \EndIf 
                  \State{generate($succ(s_i)$)} \label{supp:alg:OBATS:expansion from Open end}
                  \State lock($\Unevaluated$) \label{supp:alg:OBATS:inesrt unevaluated start}
                  \For{$s_i' \in succ(s_i)$}
                    \State{$\Unevaluated\gets{\Unevaluated\cup\{s_i'\}}$}
                    \State $\parent(s_i') \leftarrow s_i$
                  \EndFor
                  \State unlock($\Unevaluated$) \label{supp:alg:OBATS:inesrt unevaluated end}
                  \If {$succ(s_i) = \emptyset$} $Being\_Expanded \leftarrow Being\_Expanded  \backslash \{s_i\}$ \EndIf
                \Else
                  \State unlock($\open$), unlock($\Deferred$)
                \EndIf
              \EndIf
            \EndIf
            
          \EndIf
          \State {$s_i \leftarrow{NULL}$}
        \EndLoop
        \EndFor
        \caption{OBAT with SEE (OBAT$_S$), lock/unlock operations for $Being\_Expanded$ omitted for space}
        \label{supp:alg:OBATS}
    \end{algorithmic}
\end{algorithm}

Algorithm \ref{supp:alg:OBATS} shows \OBATS (OBAT with SGE).

Although the pseudocode may appear more complex than OBAT, the additional code required for \OBATS is a straightforward implementation of SGE, very similar to the previous application of SGE to PUHF and KPGBFS \cite{ShimodaF24}.

Most of the new pseudocode relative to OBAT are in the {\tt if}-block in lines \ref{supp:alg:OBATS:Unevaluated start}--\ref{supp:alg:OBATS:Unevaluated end}, which handles the parallel evaluation of states in $\Unevaluated$ by available threads.

In OBAT, when a state $s$ leaves $\open$,
the thread which removed $s$ generates $succ(s)$, the successors of $s$, and evaluates the $h$-values of $succ(s)$.

In \OBATS, when a state $s$ leaves $\open$,
the thread which removed $s$ generates $succ(s)$, but instead of evaluating $succ(s)$, inserts $succ(s)$ into $\Unevaluated$ (lines~\ref{supp:alg:OBATS:inesrt unevaluated start}-\ref{supp:alg:OBATS:inesrt unevaluated end}).
The unevaluated states in $\Unevaluated$ are evaluated and processed as soon as threads become available, having higher priority than removing states from $\Deferred$ and $\open$.

\section{Explanation of Benchmark Selection}
\label{supp:benchmark selection}

We compared GBFS, OBAT, \OBATS, KPGBFS, \KPDGBFS (KPGBFS with SGE), PUHF3, \PDUHF (PUHF3 with SGE) using a set of instances based on the Autoscale-21.11 satisficing benchmark set (42 STRIPS domains, 30 instances/domain, 1260 total instances) \cite{torralba-et-al-icaps2021}.
The Autoscale benchmarks are an improved benchmark suite based on the IPC classical planning benchmarks which were designed to compare the performance of different solvers, as advances in solvers sometimes made performance comparisons among modern solvers difficult using the classic IPC competition benchmark instances.

However, even on the Autoscale-21.11 benchmark set, there were several domains which were too easy -- all methods solved all instances, rendering these instances useless for the purpose of comparing coverage among the methods.

Therefore, for domains where  (1) all methods solved all instances for $k=4$, $k=8$, and $k=16$ threads, and (2) a parameterized instance generator for the domain is available in the Autoscale repository, we replaced the Autoscale-21.11 instances with more difficult instances generated using the same Autoscale instance generator.
Specifically, the domains where criteria (1) and (2) above applied were \pddl{gripper}, and \pddl{miconic}.

\begin{itemize}
\item \pddl{gripper}:
  The Autoscale-21.11 set used values of parameter $n$ from 20 to 165, in increments of 5 (30 instances).
  Our instances used $n$ from 175 to 465, in increments of 10 (30 instances).

  \item \pddl{miconic}:

    The Autoscale \pddl{miconic} instances are generated using two parameters,
    $\textit{passengers}$ and  $\textit{floors}$.
    In the Autoscale-21.11 set (30 instances), $19 \leq \textit{passengers} \leq 155$, where $\textit{passengers}$ monotonically increased by 4 or 5 between consecutive instances,  and
    $11 \leq \textit{floors} \leq 124$, monotonically increasing in increments of 3 or 4.

    Similarly, our replacement instances were generated using the following procedure, which calls the Autoscale {\tt generate\_instance} function for \pddl{miconic}:
    \begin{small}
    \begin{verbatim}
passengers=155, floors=124
for i in range(0,90):
     if(i%
          passengers+=4
     else:
          passengers+=5
     floors+=4
     if(i%
          generate_instance(passengers,floors)
\end{verbatim}
\end{small}
    
\end{itemize}

The sas files are available at \url{https://github.com/TakuShimoda/AAAI25}.

\clearpage

\section{Comparisons Without Separate Generation and Evaluation (SGE)}

\begin{figure}[H]
  \centering
  \begin{subfigure}[]{0.18\columnwidth}
    \includegraphics[width=\textwidth,trim={1cm 0.4cm 1cm 0.1cm},,clip]{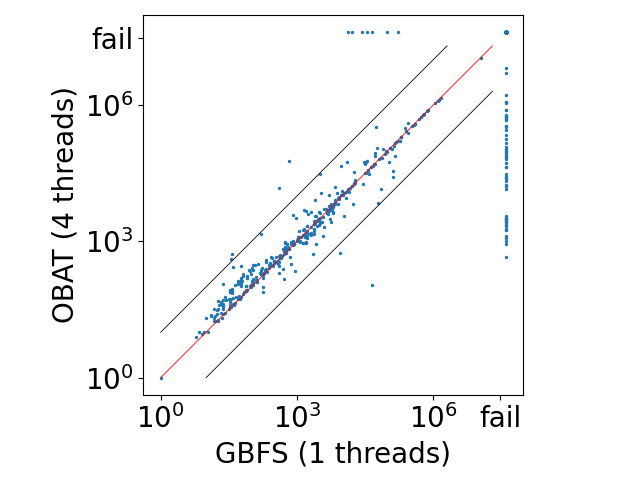}
    \caption{}
  \end{subfigure}
  \begin{subfigure}[]{0.18\columnwidth}
    \includegraphics[width=\textwidth,trim={1cm 0.4cm 1cm 0.1cm},,clip]{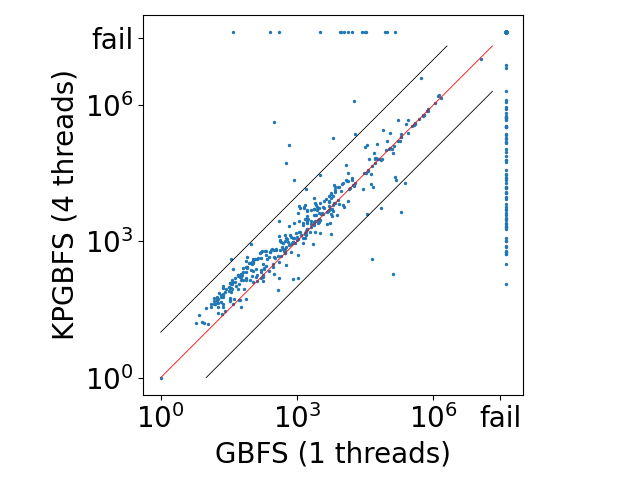}
    \caption{}
  \end{subfigure}
  \begin{subfigure}[]{0.18\columnwidth}
    \includegraphics[width=\textwidth,trim={1cm 0.4cm 1cm 0.1cm},,clip]{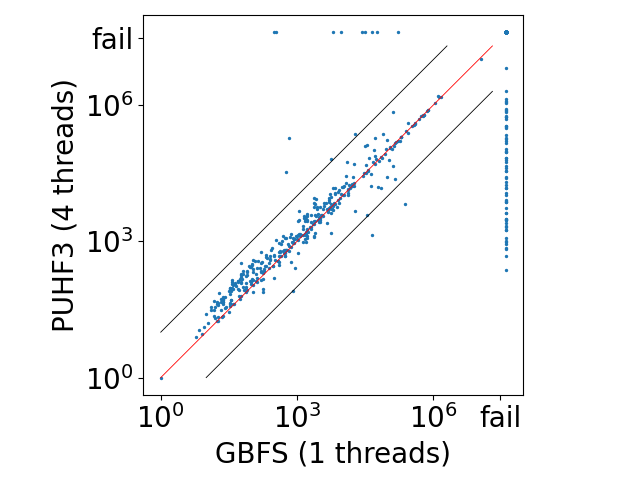}
    \caption{}
  \end{subfigure}
  \begin{subfigure}[]{0.18\columnwidth}
    \includegraphics[width=\textwidth,trim={1cm 0.4cm 1cm 0.1cm},,clip]{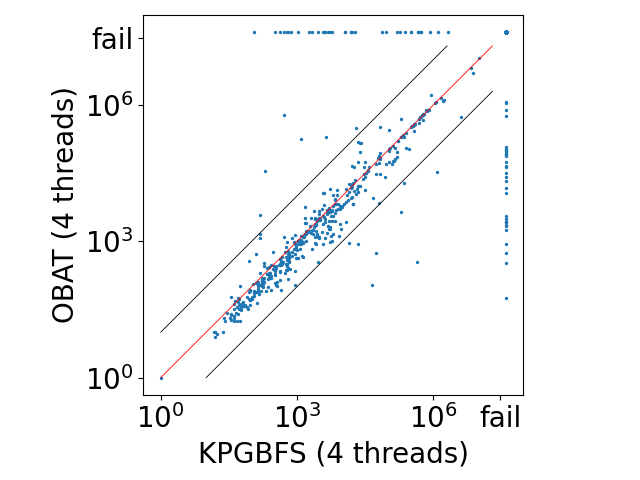}
    \caption{}
  \end{subfigure}
  \begin{subfigure}[]{0.18\columnwidth}
    \includegraphics[width=\textwidth,trim={1cm 0.4cm 1cm 0.1cm},,clip]{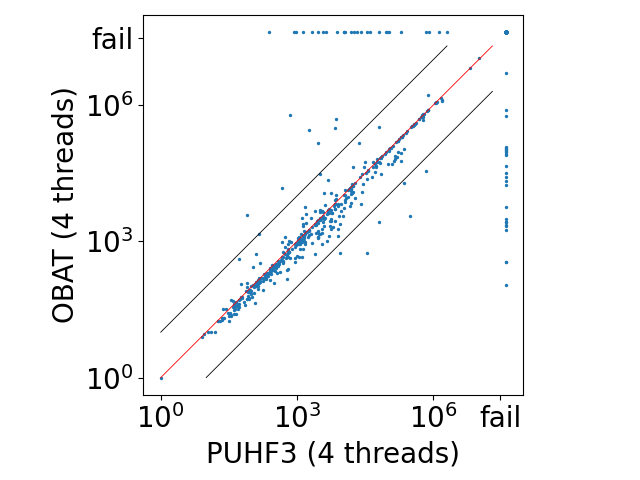}
    \caption{}
  \end{subfigure}

  \begin{subfigure}[]{0.18\columnwidth}
    \includegraphics[width=\textwidth,trim={1cm 0.4cm 1cm 0.1cm},,clip]{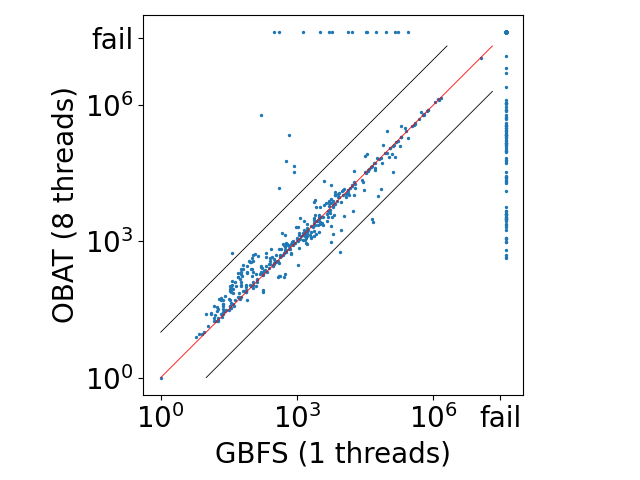}
    \caption{}
  \end{subfigure}
  \begin{subfigure}[]{0.18\columnwidth}
    \includegraphics[width=\textwidth,trim={1cm 0.4cm 1cm 0.1cm},,clip]{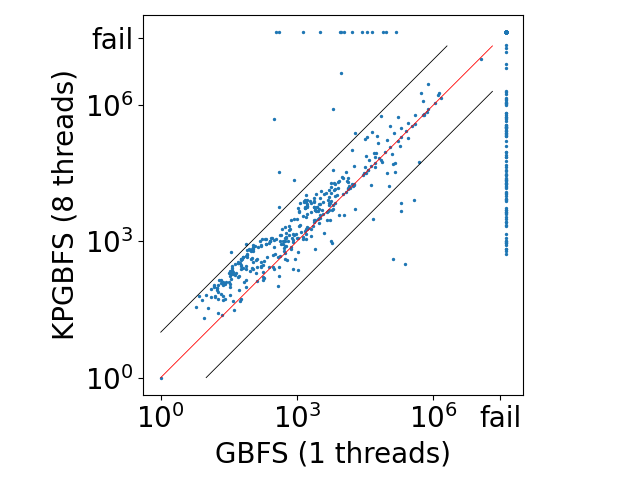}
    \caption{}
  \end{subfigure}
  \begin{subfigure}[]{0.18\columnwidth}
    \includegraphics[width=\textwidth,trim={1cm 0.4cm 1cm 0.1cm},,clip]{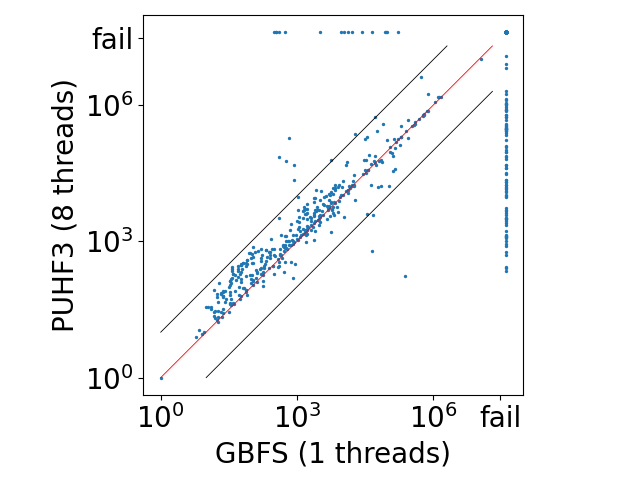}
    \caption{}
  \end{subfigure}
  \begin{subfigure}[]{0.18\columnwidth}
    \includegraphics[width=\textwidth,trim={1cm 0.4cm 1cm 0.1cm},,clip]{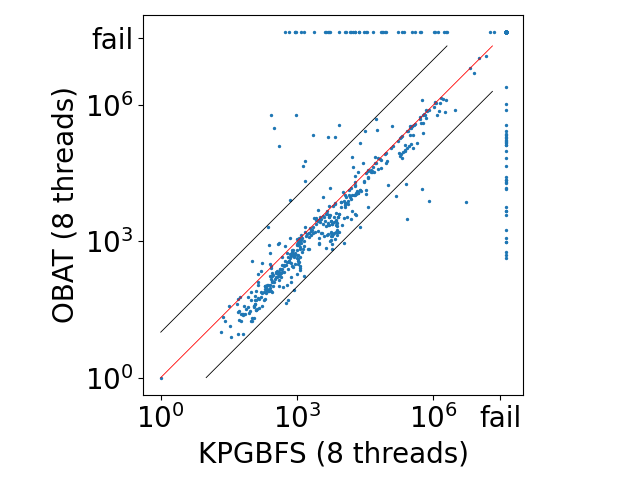}
    \caption{}
  \end{subfigure}
  \begin{subfigure}[]{0.18\columnwidth}
    \includegraphics[width=\textwidth,trim={1cm 0.4cm 1cm 0.1cm},,clip]{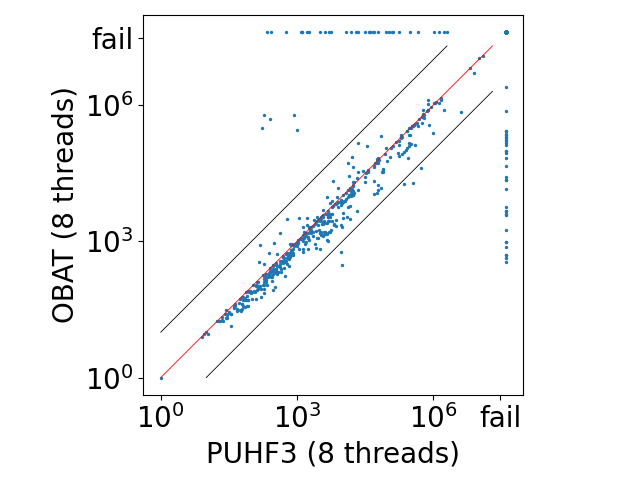}
    \caption{}
  \end{subfigure}

  \begin{subfigure}[]{0.18\columnwidth}
    \includegraphics[width=\textwidth,trim={1cm 0.4cm 1cm 0.1cm},,clip]{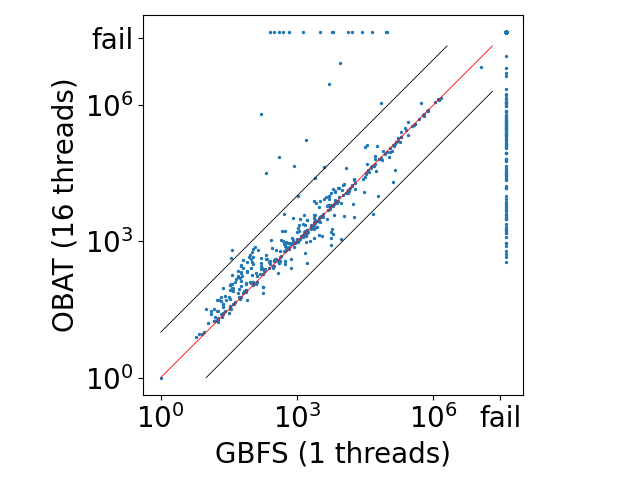}
    \caption{}
  \end{subfigure}
  \begin{subfigure}[]{0.18\columnwidth}
    \includegraphics[width=\textwidth,trim={1cm 0.4cm 1cm 0.1cm},,clip]{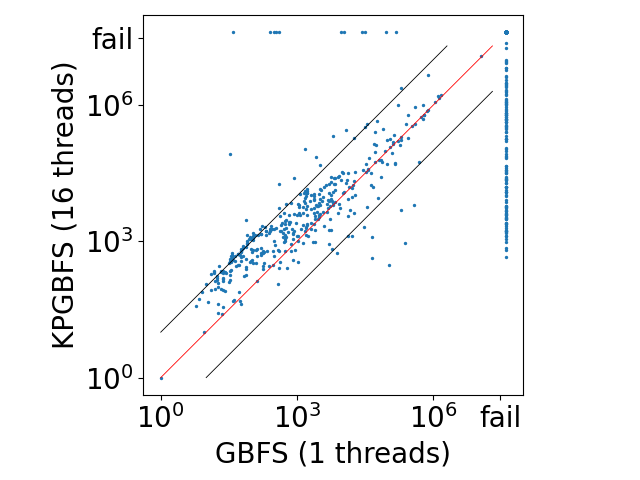}
    \caption{}
  \end{subfigure}
  \begin{subfigure}[]{0.18\columnwidth}
    \includegraphics[width=\textwidth,trim={1cm 0.4cm 1cm 0.1cm},,clip]{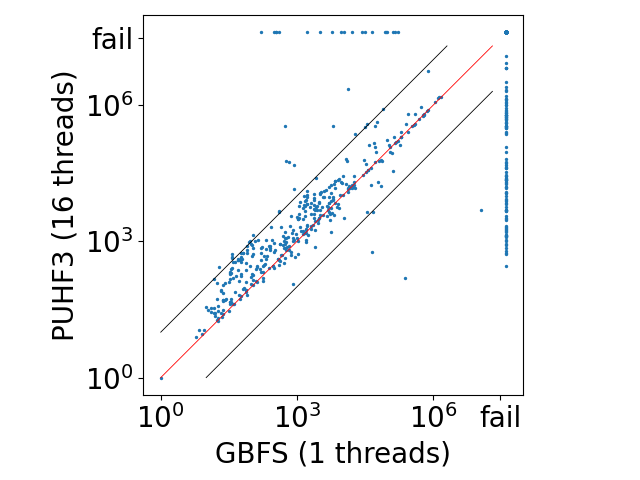}
    \caption{}
  \end{subfigure}
  \begin{subfigure}[]{0.18\columnwidth}
    \includegraphics[width=\textwidth,trim={1cm 0.4cm 1cm 0.1cm},,clip]{figures/expansions/16threads/expansions_OBAT_K=16_vs_KPGBFS_K=16.png}
    \caption{}
  \end{subfigure}
  \begin{subfigure}[]{0.18\columnwidth}
    \includegraphics[width=\textwidth,trim={1cm 0.4cm 1cm 0.1cm},,clip]{figures/expansions/16threads/expansions_OBAT_K=16_vs_PUHF3_K=16.png}
    \caption{}
  \end{subfigure}

  \centering
  \caption{ Number of states expanded, Diagonal lines are $y=0.1x$, $y=x$, and $y=10x$}
  \label{supp:fig:expansions-comparisons-OBAT}
  \centering
\end{figure}

\begin{figure}[H]
  \centering
  \begin{subfigure}[]{0.18\columnwidth}
    \includegraphics[width=\textwidth,trim={1cm 0.4cm 1cm 0.1cm},,clip]{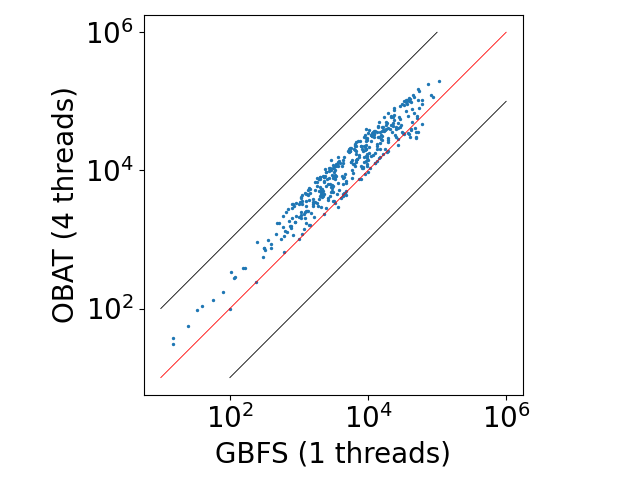}
    \caption{}
  \end{subfigure}
  \begin{subfigure}[]{0.18\columnwidth}
    \includegraphics[width=\textwidth,trim={1cm 0.4cm 1cm 0.1cm},,clip]{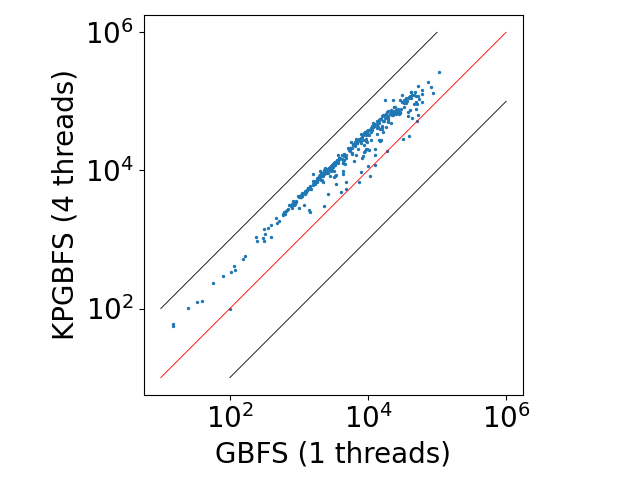}
    \caption{}
  \end{subfigure}
  \begin{subfigure}[]{0.18\columnwidth}
    \includegraphics[width=\textwidth,trim={1cm 0.4cm 1cm 0.1cm},,clip]{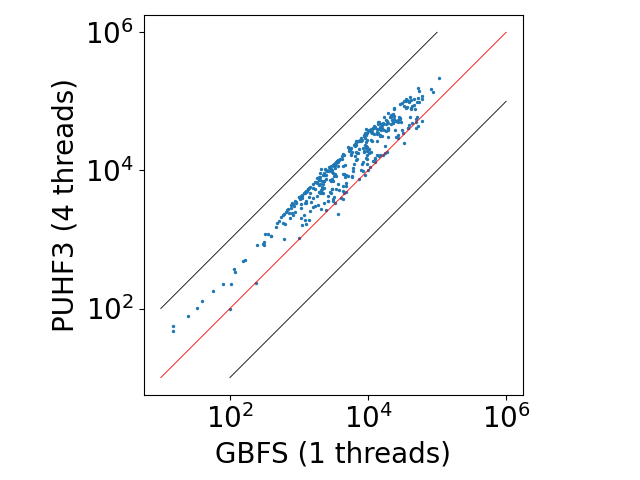}
    \caption{}
  \end{subfigure}
  \begin{subfigure}[]{0.18\columnwidth}
    \includegraphics[width=\textwidth,trim={1cm 0.4cm 1cm 0.1cm},,clip]{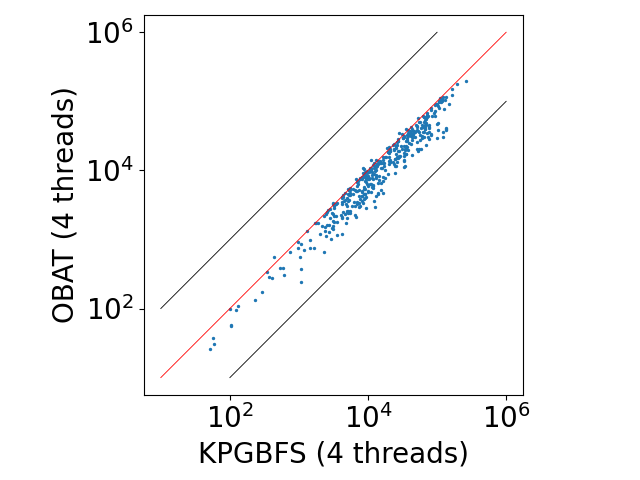}
    \caption{}
  \end{subfigure}
  \begin{subfigure}[]{0.18\columnwidth}
    \includegraphics[width=\textwidth,trim={1cm 0.4cm 1cm 0.1cm},,clip]{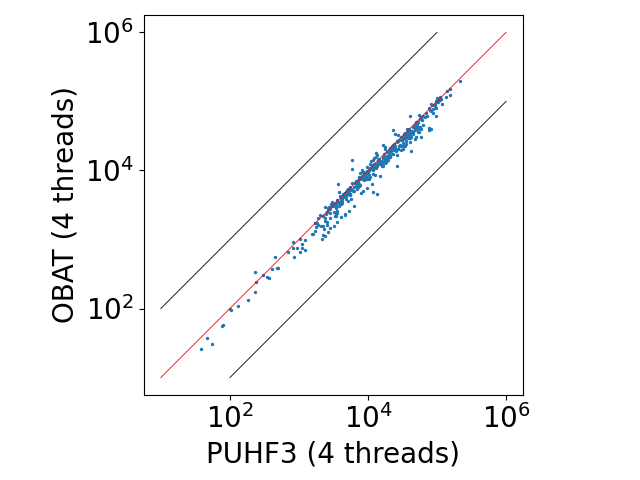}
    \caption{}
  \end{subfigure}

  \begin{subfigure}[]{0.18\columnwidth}
    \includegraphics[width=\textwidth,trim={1cm 0.4cm 1cm 0.1cm},,clip]{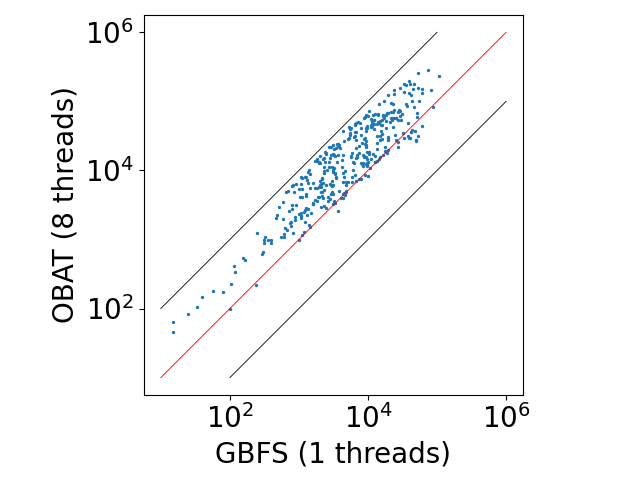}
    \caption{}
  \end{subfigure}
  \begin{subfigure}[]{0.18\columnwidth}
    \includegraphics[width=\textwidth,trim={1cm 0.4cm 1cm 0.1cm},,clip]{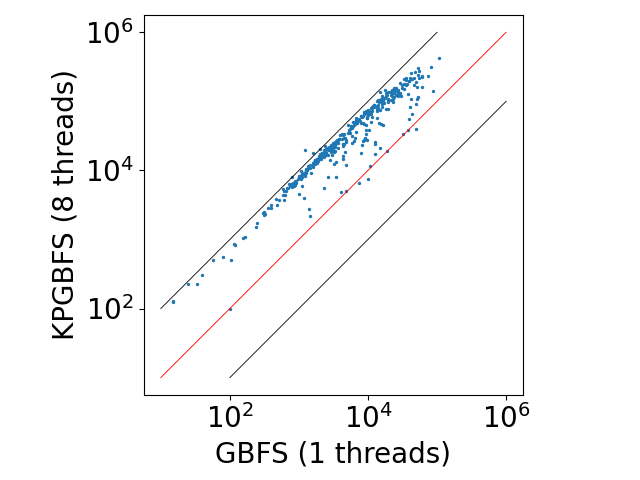}
    \caption{}
  \end{subfigure}
  \begin{subfigure}[]{0.18\columnwidth}
    \includegraphics[width=\textwidth,trim={1cm 0.4cm 1cm 0.1cm},,clip]{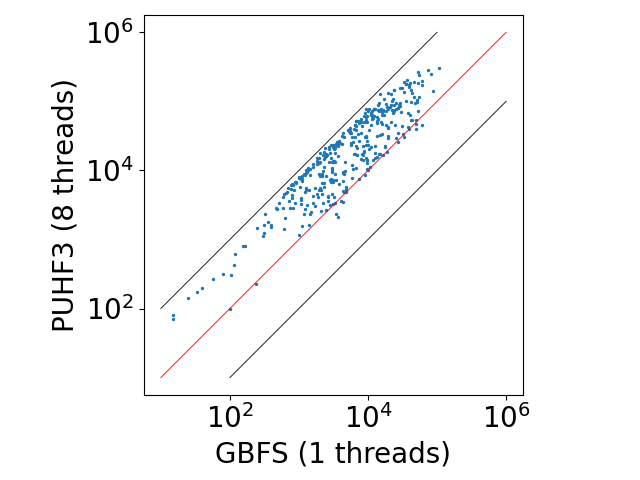}
    \caption{}
  \end{subfigure}
  \begin{subfigure}[]{0.18\columnwidth}
    \includegraphics[width=\textwidth,trim={1cm 0.4cm 1cm 0.1cm},,clip]{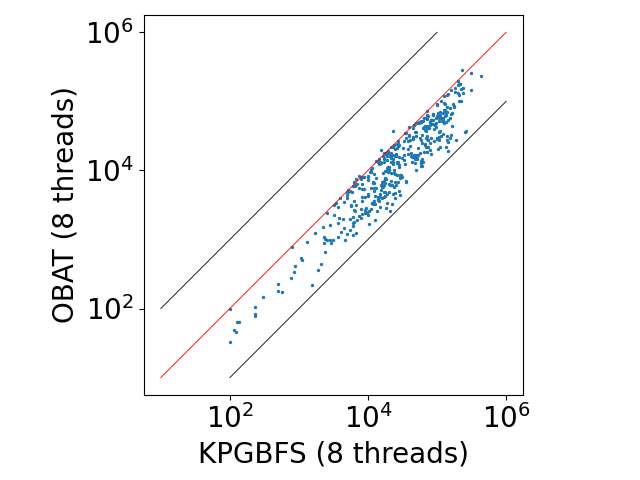}
    \caption{}
  \end{subfigure}
  \begin{subfigure}[]{0.18\columnwidth}
    \includegraphics[width=\textwidth,trim={1cm 0.4cm 1cm 0.1cm},,clip]{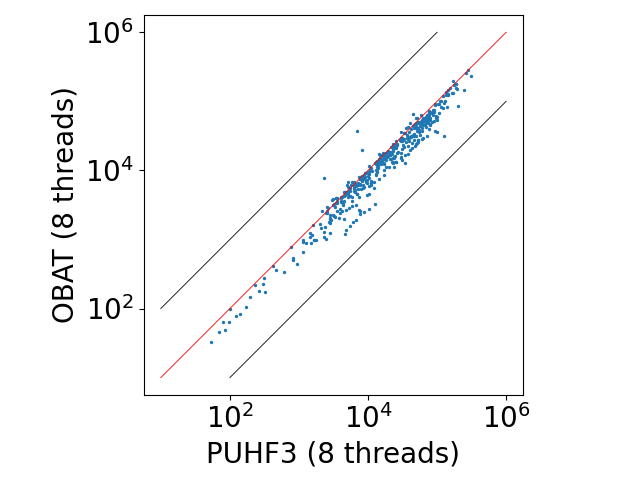}
    \caption{}
  \end{subfigure}

  \begin{subfigure}[]{0.18\columnwidth}
    \includegraphics[width=\textwidth,trim={1cm 0.4cm 1cm 0.1cm},,clip]{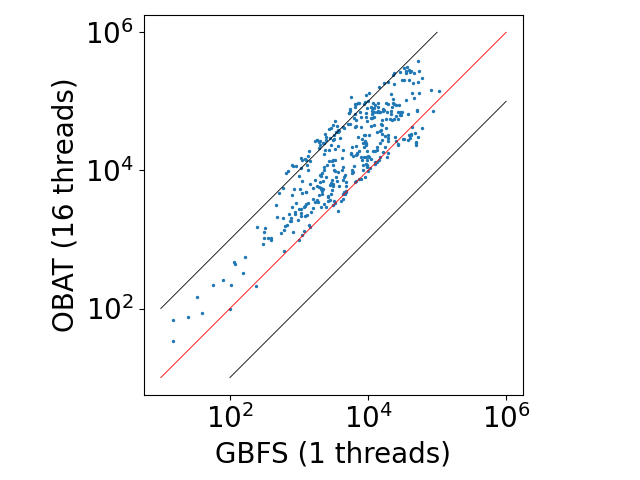}
    \caption{}
  \end{subfigure}
  \begin{subfigure}[]{0.18\columnwidth}
    \includegraphics[width=\textwidth,trim={1cm 0.4cm 1cm 0.1cm},,clip]{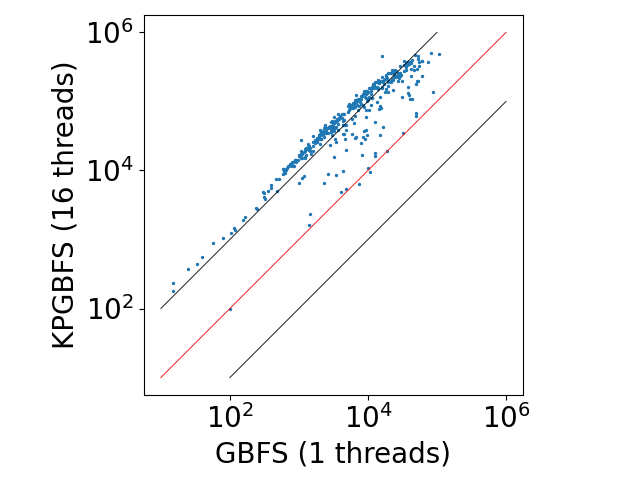}
    \caption{}
  \end{subfigure}
  \begin{subfigure}[]{0.18\columnwidth}
    \includegraphics[width=\textwidth,trim={1cm 0.4cm 1cm 0.1cm},,clip]{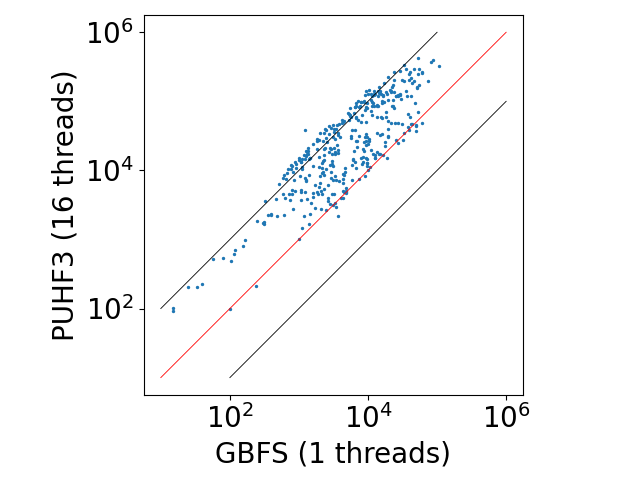}
    \caption{}
  \end{subfigure}
  \begin{subfigure}[]{0.18\columnwidth}
    \includegraphics[width=\textwidth,trim={1cm 0.4cm 1cm 0.1cm},,clip]{figures/evaluation_rate/16threads/evaluation_rate_OBAT_K=16_vs_KPGBFS_K=16.png}
    \caption{}
  \end{subfigure}
  \begin{subfigure}[]{0.18\columnwidth}
    \includegraphics[width=\textwidth,trim={1cm 0.4cm 1cm 0.1cm},,clip]{figures/evaluation_rate/16threads/evaluation_rate_OBAT_K=16_vs_PUHF3_K=16.png}
    \caption{}
  \end{subfigure}

  \centering
  \caption{ State evaluation rate comparison (states/second), Diagonal lines are $y=0.1x$, $y=x$, and $y=10x$}
  \label{supp:fig:evaluation-rate-comparisons-OBAT}
  
  \centering
\end{figure}

\begin{figure}[H]
  \centering
  \begin{subfigure}[]{0.18\columnwidth}
    \includegraphics[width=\textwidth,trim={1cm 0.4cm 1cm 0.1cm},,clip]{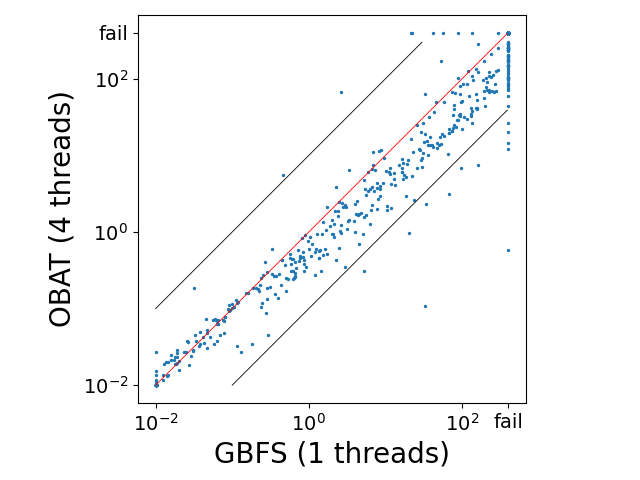}
    \caption{}
  \end{subfigure}
  \begin{subfigure}[]{0.18\columnwidth}
    \includegraphics[width=\textwidth,trim={1cm 0.4cm 1cm 0.1cm},,clip]{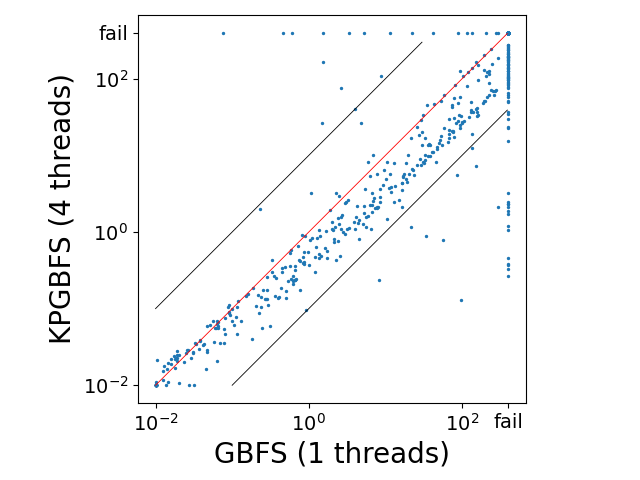}
    \caption{}
  \end{subfigure}
  \begin{subfigure}[]{0.18\columnwidth}
    \includegraphics[width=\textwidth,trim={1cm 0.4cm 1cm 0.1cm},,clip]{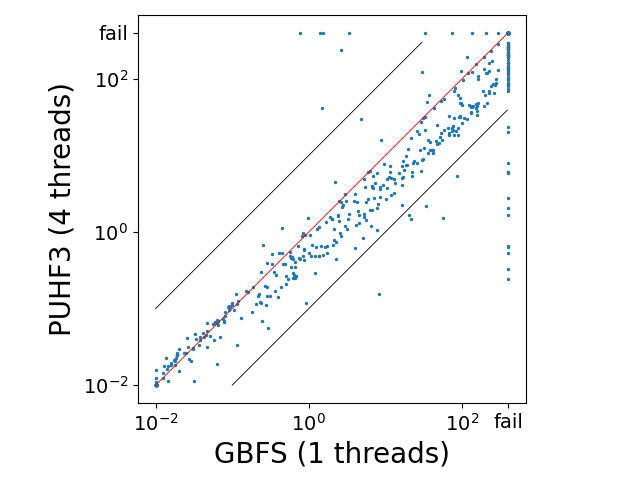}
    \caption{}
  \end{subfigure}
  \begin{subfigure}[]{0.18\columnwidth}
    \includegraphics[width=\textwidth,trim={1cm 0.4cm 1cm 0.1cm},,clip]{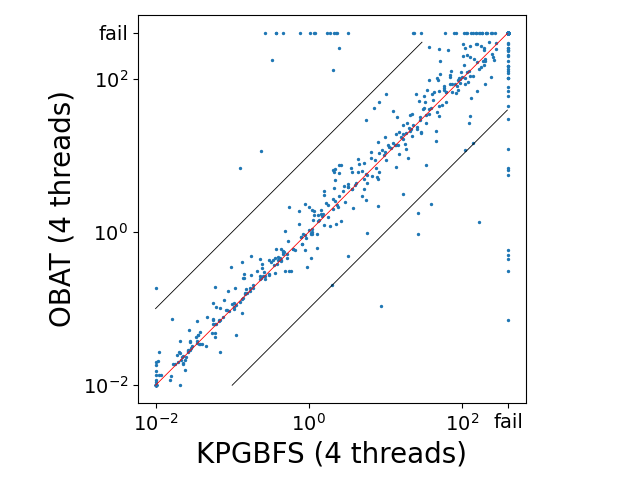}
    \caption{}
  \end{subfigure}
  \begin{subfigure}[]{0.18\columnwidth}
    \includegraphics[width=\textwidth,trim={1cm 0.4cm 1cm 0.1cm},,clip]{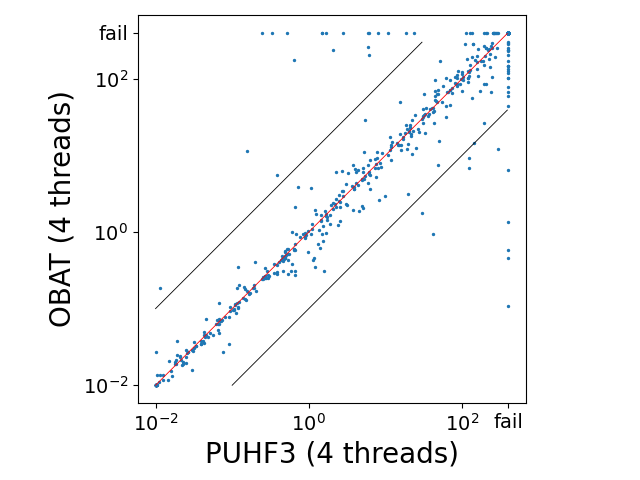}
    \caption{}
  \end{subfigure}

  \begin{subfigure}[]{0.18\columnwidth}
    \includegraphics[width=\textwidth,trim={1cm 0.4cm 1cm 0.1cm},,clip]{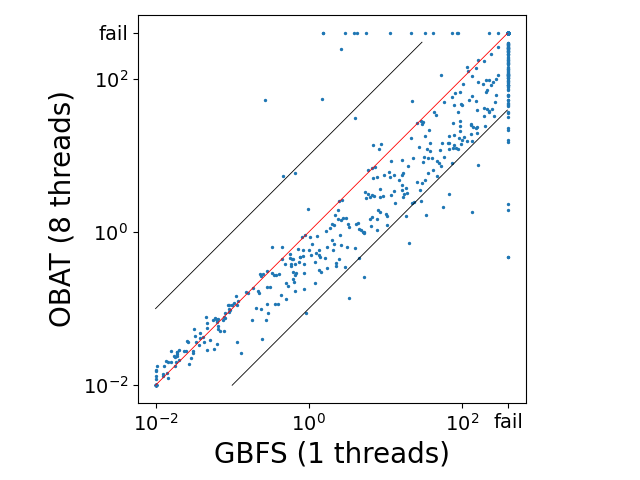}
    \caption{}
  \end{subfigure}
  \begin{subfigure}[]{0.18\columnwidth}
    \includegraphics[width=\textwidth,trim={1cm 0.4cm 1cm 0.1cm},,clip]{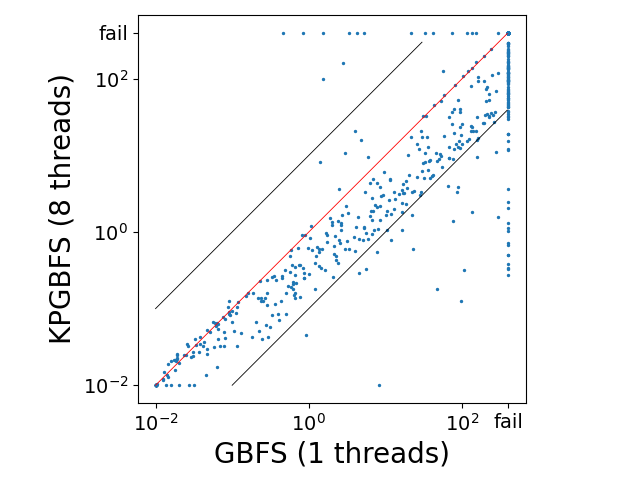}
    \caption{}
  \end{subfigure}
  \begin{subfigure}[]{0.18\columnwidth}
    \includegraphics[width=\textwidth,trim={1cm 0.4cm 1cm 0.1cm},,clip]{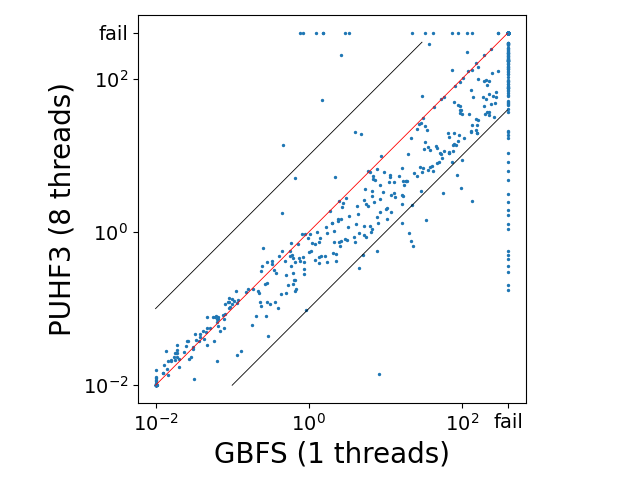}
    \caption{}
  \end{subfigure}
  \begin{subfigure}[]{0.18\columnwidth}
    \includegraphics[width=\textwidth,trim={1cm 0.4cm 1cm 0.1cm},,clip]{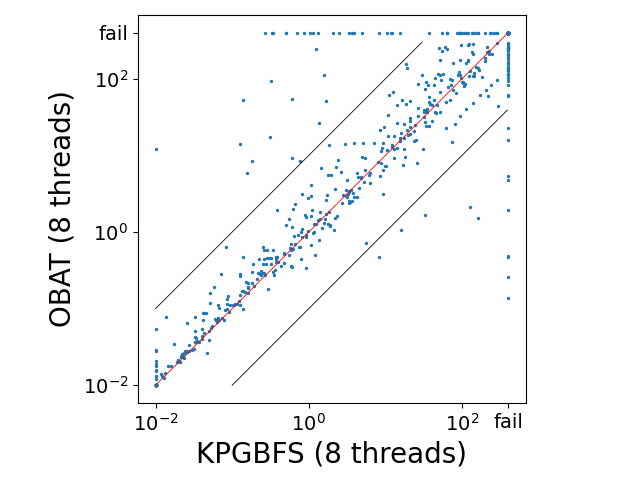}
    \caption{}
  \end{subfigure}
  \begin{subfigure}[]{0.18\columnwidth}
    \includegraphics[width=\textwidth,trim={1cm 0.4cm 1cm 0.1cm},,clip]{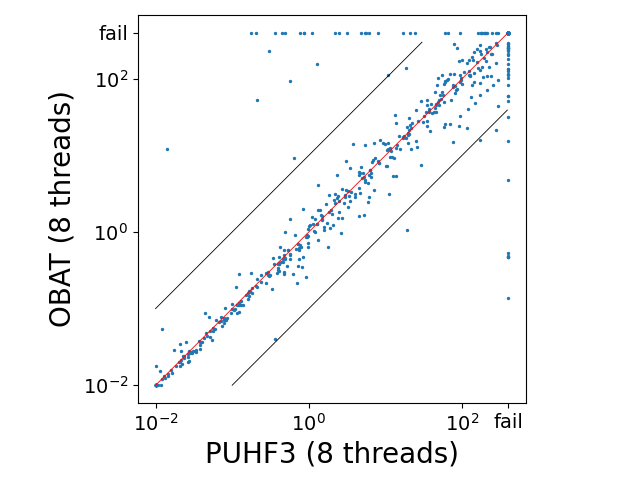}
    \caption{}
  \end{subfigure}

  \begin{subfigure}[]{0.18\columnwidth}
    \includegraphics[width=\textwidth,trim={1cm 0.4cm 1cm 0.1cm},,clip]{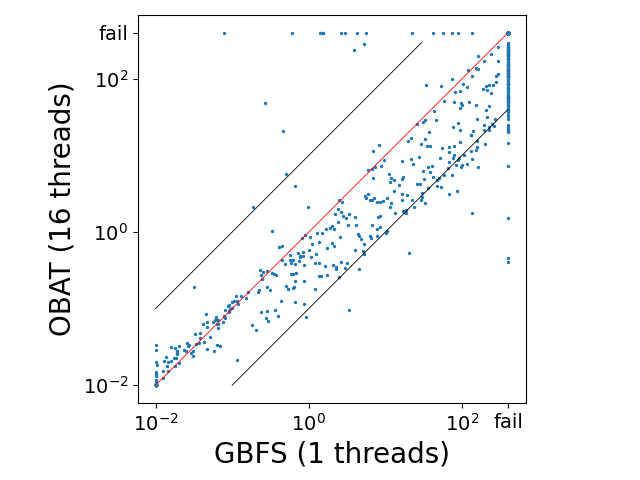}
    \caption{}
  \end{subfigure}
  \begin{subfigure}[]{0.18\columnwidth}
    \includegraphics[width=\textwidth,trim={1cm 0.4cm 1cm 0.1cm},,clip]{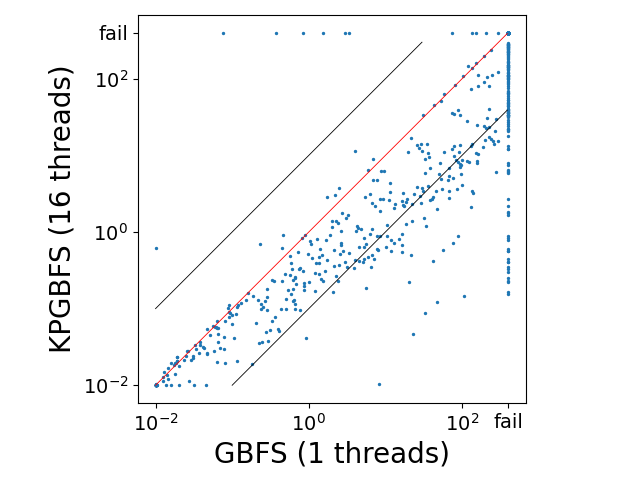}
    \caption{}
  \end{subfigure}
  \begin{subfigure}[]{0.18\columnwidth}
    \includegraphics[width=\textwidth,trim={1cm 0.4cm 1cm 0.1cm},,clip]{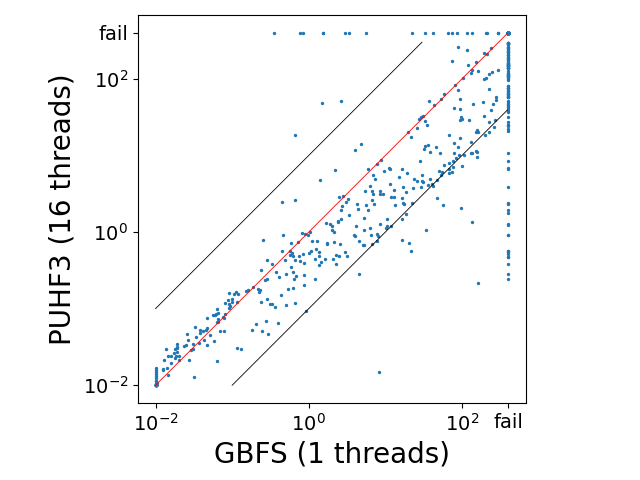}
    \caption{}
  \end{subfigure}
  \begin{subfigure}[]{0.18\columnwidth}
    \includegraphics[width=\textwidth,trim={1cm 0.4cm 1cm 0.1cm},,clip]{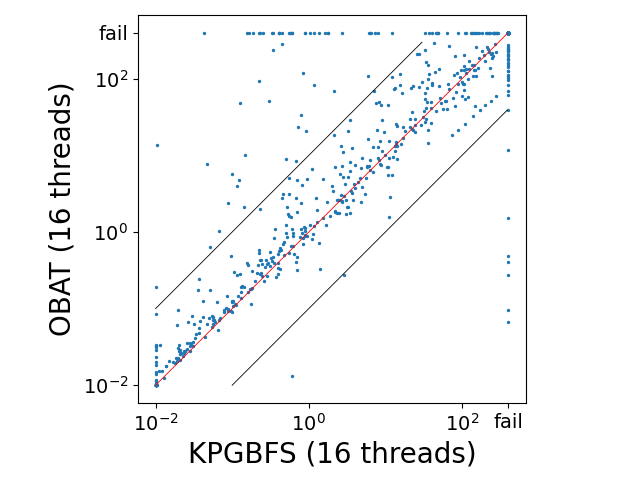}
    \caption{}
  \end{subfigure}
  \begin{subfigure}[]{0.18\columnwidth}
    \includegraphics[width=\textwidth,trim={1cm 0.4cm 1cm 0.1cm},,clip]{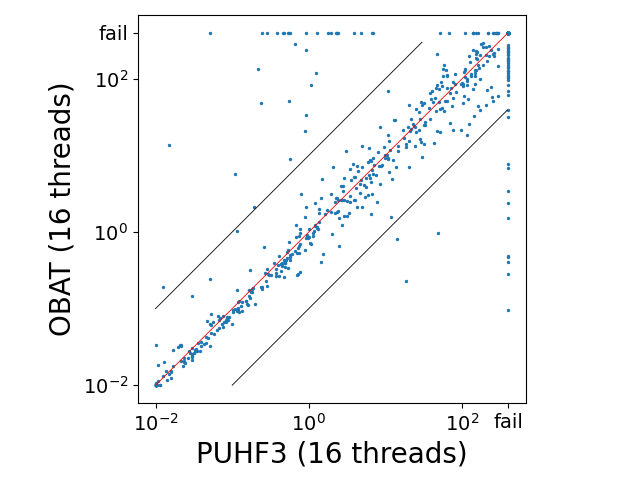}
    \caption{}
  \end{subfigure}

  \centering
  \caption{ Search time (seconds) ``fail''= out of time/memory, diagonal lines are $y=0.1x$, $y=x$, and $y=10x$}
  \label{supp:fig:search-time-comparisons-OBAT}
 
  \centering
\end{figure}

\section{ Comparisons Including Separate Generation and Evaluation (SGE)}
\begin{figure}[H]
  \centering
  \begin{subfigure}[]{0.18\columnwidth}
    \includegraphics[width=\textwidth,trim={1cm 0.4cm 1cm 0.1cm},,clip]{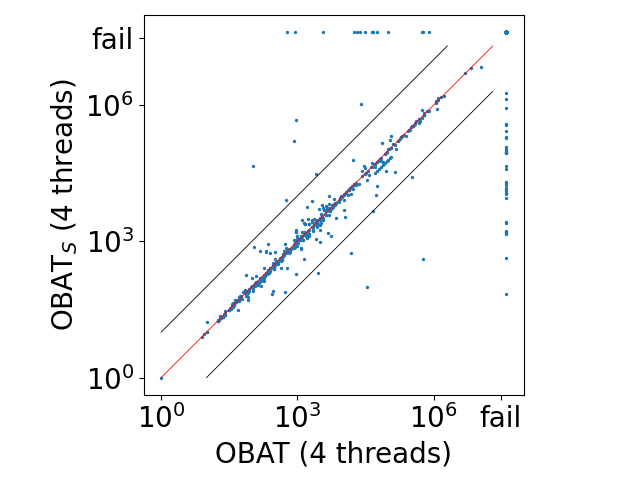}
    \caption{}
  \end{subfigure}
  \begin{subfigure}[]{0.18\columnwidth}
    \includegraphics[width=\textwidth,trim={1cm 0.4cm 1cm 0.1cm},,clip]{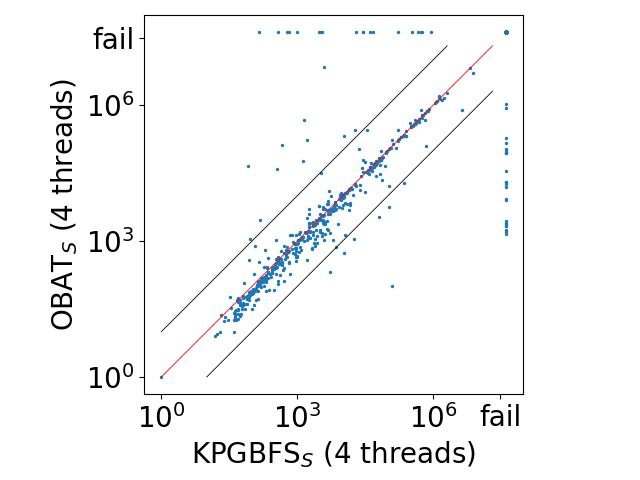}
    \caption{}
  \end{subfigure}
  \begin{subfigure}[]{0.18\columnwidth}
    \includegraphics[width=\textwidth,trim={1cm 0.4cm 1cm 0.1cm},,clip]{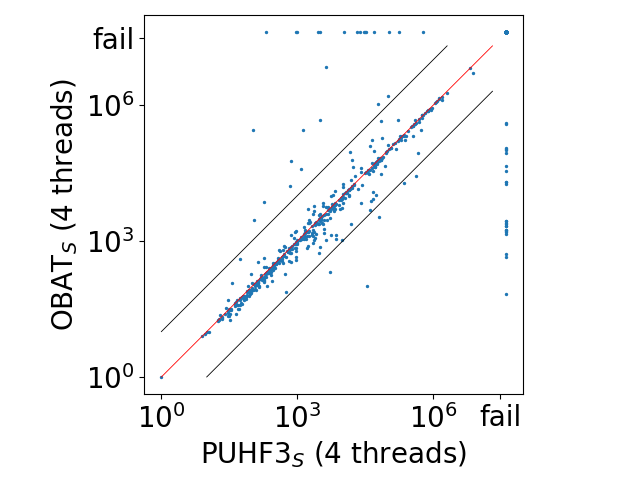}
    \caption{}
  \end{subfigure}
  \begin{subfigure}[]{0.18\columnwidth}
    \includegraphics[width=\textwidth,trim={1cm 0.4cm 1cm 0.1cm},,clip]{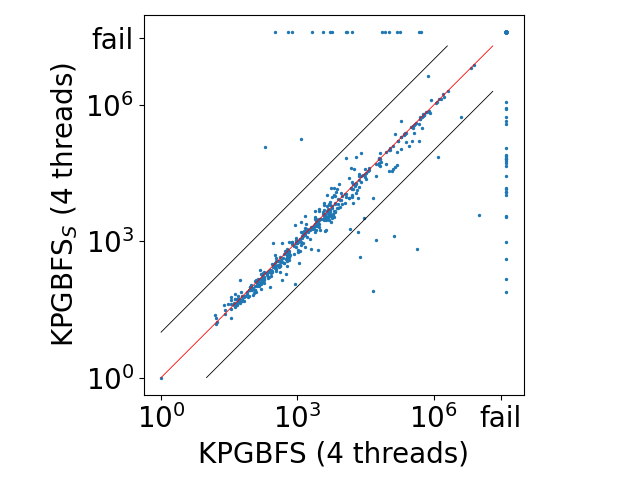}
    \caption{}
  \end{subfigure}
  \begin{subfigure}[]{0.18\columnwidth}
    \includegraphics[width=\textwidth,trim={1cm 0.4cm 1cm 0.1cm},,clip]{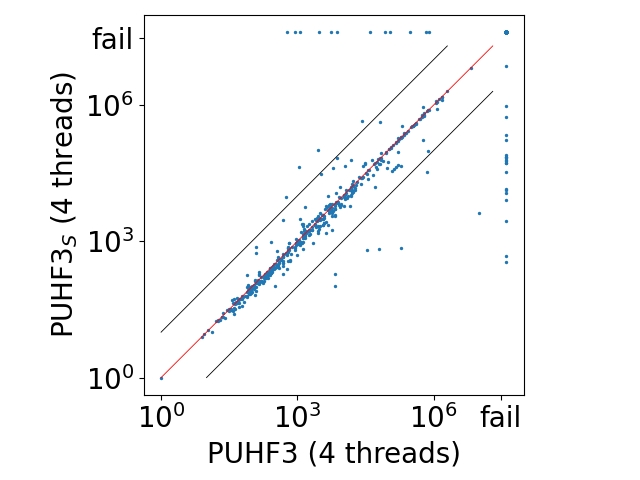}
    \caption{}
  \end{subfigure}

  \begin{subfigure}[]{0.18\columnwidth}
    \includegraphics[width=\textwidth,trim={1cm 0.4cm 1cm 0.1cm},,clip]{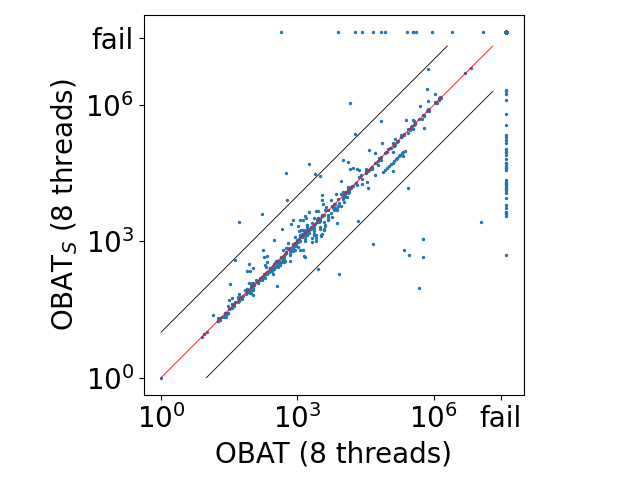}
    \caption{}
  \end{subfigure}
  \begin{subfigure}[]{0.18\columnwidth}
    \includegraphics[width=\textwidth,trim={1cm 0.4cm 1cm 0.1cm},,clip]{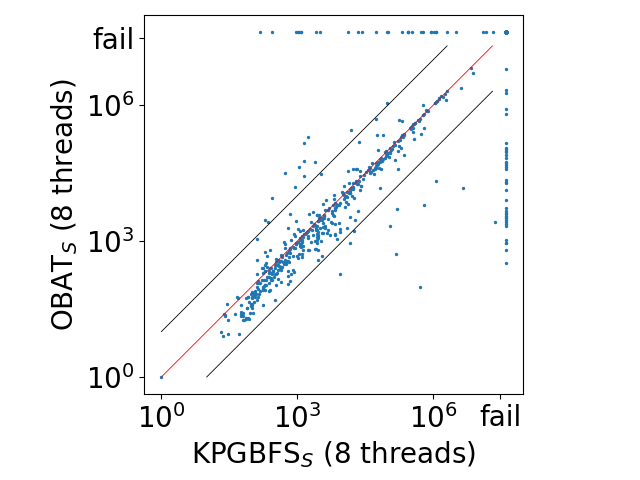}
    \caption{}
  \end{subfigure}
  \begin{subfigure}[]{0.18\columnwidth}
    \includegraphics[width=\textwidth,trim={1cm 0.4cm 1cm 0.1cm},,clip]{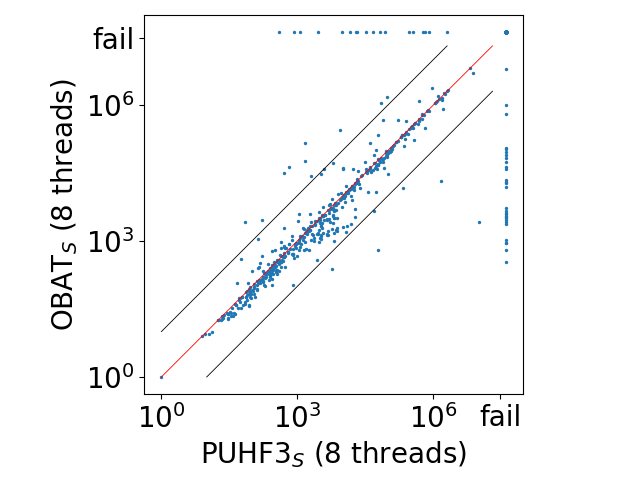}
    \caption{}
  \end{subfigure}
  \begin{subfigure}[]{0.18\columnwidth}
    \includegraphics[width=\textwidth,trim={1cm 0.4cm 1cm 0.1cm},,clip]{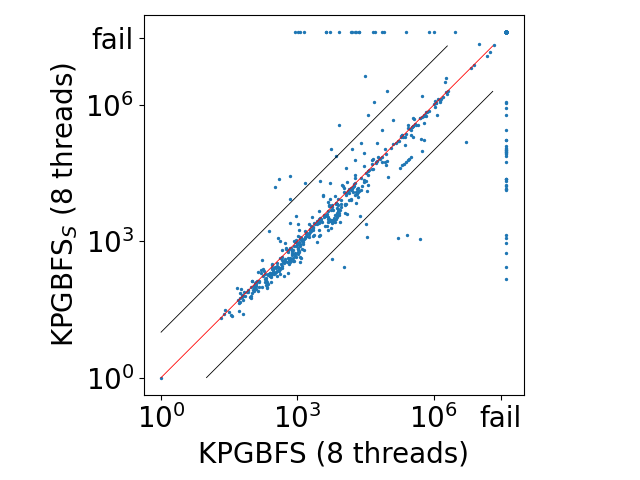}
    \caption{}
  \end{subfigure}
  \begin{subfigure}[]{0.18\columnwidth}
    \includegraphics[width=\textwidth,trim={1cm 0.4cm 1cm 0.1cm},,clip]{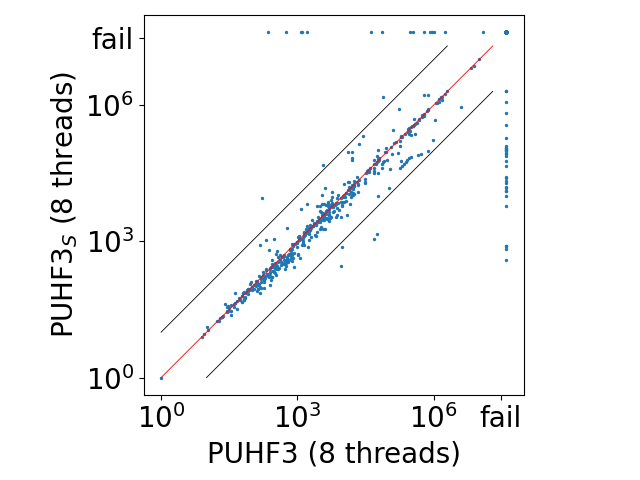}
    \caption{}
  \end{subfigure}

  \begin{subfigure}[]{0.18\columnwidth}
    \includegraphics[width=\textwidth,trim={1cm 0.4cm 1cm 0.1cm},,clip]{figures/expansions/16threads/expansions_OBAT_S_K=16_vs_OBAT_K=16.png}
    \caption{}
  \end{subfigure}
  \begin{subfigure}[]{0.18\columnwidth}
    \includegraphics[width=\textwidth,trim={1cm 0.4cm 1cm 0.1cm},,clip]{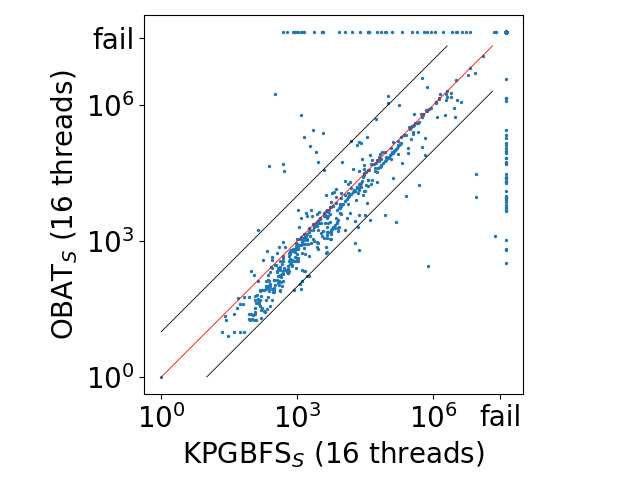}
    \caption{}
  \end{subfigure}
  \begin{subfigure}[]{0.18\columnwidth}
    \includegraphics[width=\textwidth,trim={1cm 0.4cm 1cm 0.1cm},,clip]{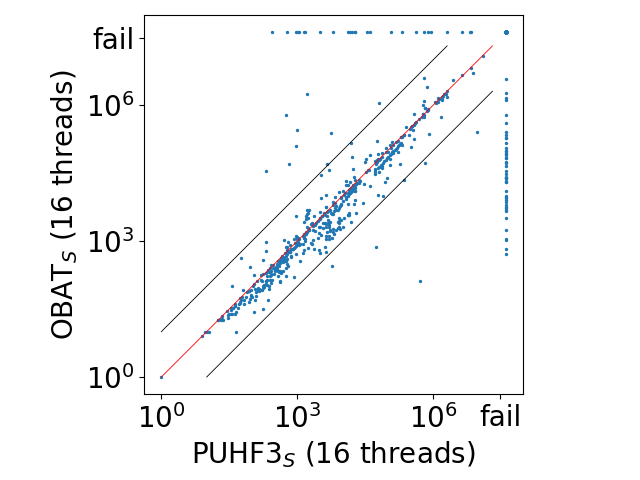}
    \caption{}
  \end{subfigure}
  \begin{subfigure}[]{0.18\columnwidth}
    \includegraphics[width=\textwidth,trim={1cm 0.4cm 1cm 0.1cm},,clip]{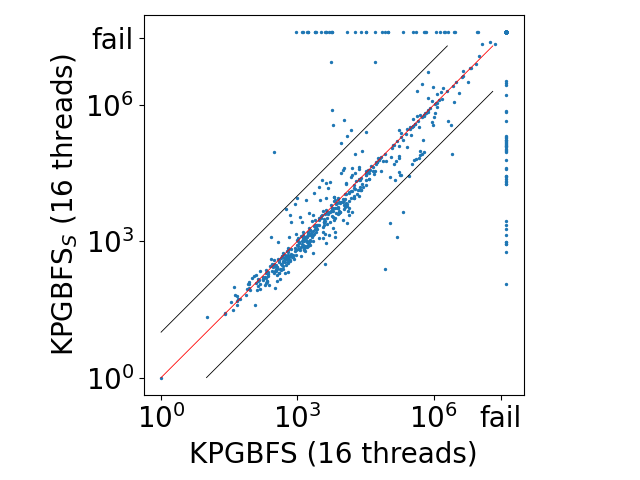}
    \caption{}
  \end{subfigure}
  \begin{subfigure}[]{0.18\columnwidth}
    \includegraphics[width=\textwidth,trim={1cm 0.4cm 1cm 0.1cm},,clip]{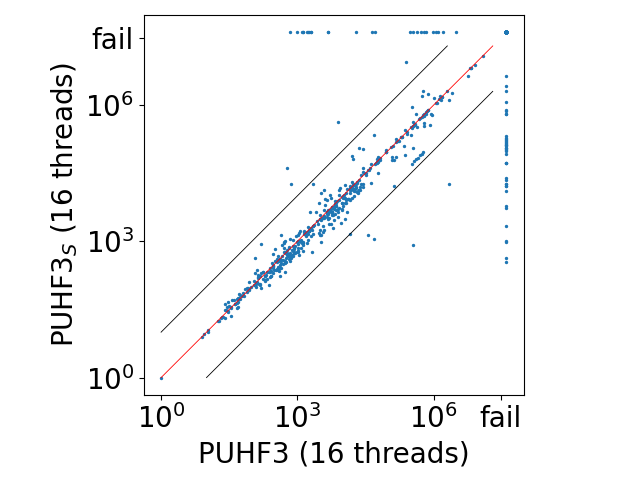}
    \caption{}
  \end{subfigure}

  \caption{ Number of states expanded, Diagonal lines are $y=0.1x$, $y=x$, and $y=10x$}
  \label{supp:fig:expansions-comparisons-OBAT_S}

\centering
\end{figure}

\begin{figure}[H]
  \centering
  \begin{subfigure}[]{0.18\columnwidth}
    \includegraphics[width=\textwidth,trim={1cm 0.4cm 1cm 0.1cm},,clip]{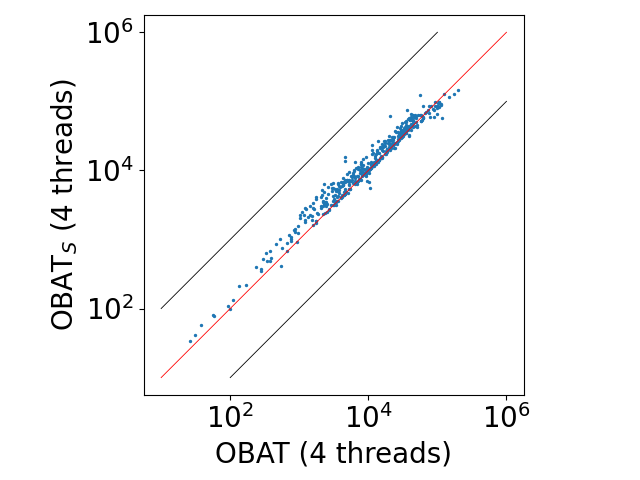}
    \caption{}
  \end{subfigure}
  \begin{subfigure}[]{0.18\columnwidth}
    \includegraphics[width=\textwidth,trim={1cm 0.4cm 1cm 0.1cm},,clip]{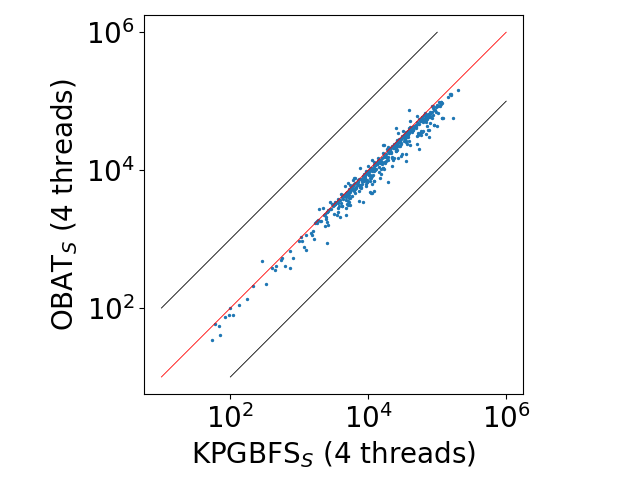}
    \caption{}
  \end{subfigure}
  \begin{subfigure}[]{0.18\columnwidth}
    \includegraphics[width=\textwidth,trim={1cm 0.4cm 1cm 0.1cm},,clip]{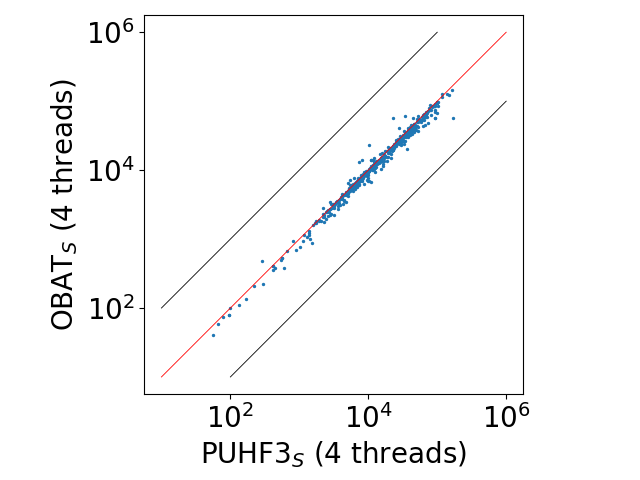}
    \caption{}
  \end{subfigure}
  \begin{subfigure}[]{0.18\columnwidth}
    \includegraphics[width=\textwidth,trim={1cm 0.4cm 1cm 0.1cm},,clip]{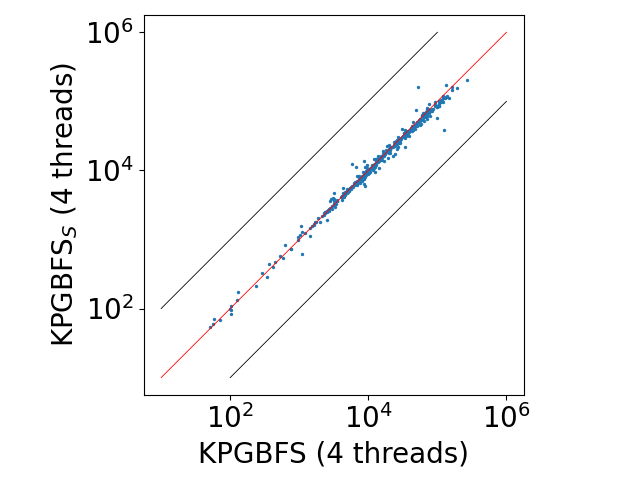}
    \caption{}
  \end{subfigure}
  \begin{subfigure}[]{0.18\columnwidth}
    \includegraphics[width=\textwidth,trim={1cm 0.4cm 1cm 0.1cm},,clip]{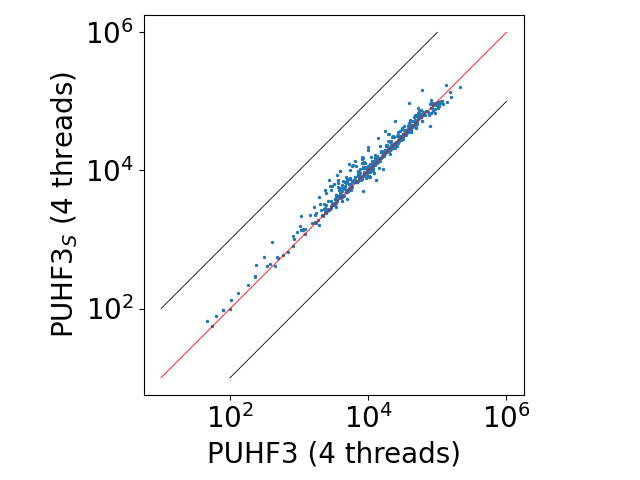}
    \caption{}
  \end{subfigure}

  \begin{subfigure}[]{0.18\columnwidth}
    \includegraphics[width=\textwidth,trim={1cm 0.4cm 1cm 0.1cm},,clip]{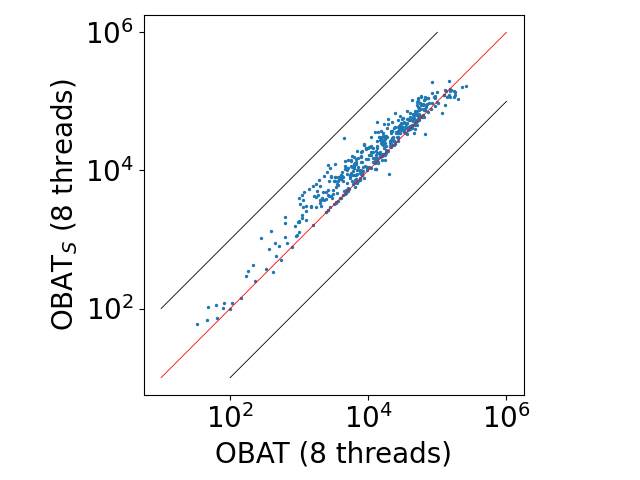}
    \caption{}
  \end{subfigure}
  \begin{subfigure}[]{0.18\columnwidth}
    \includegraphics[width=\textwidth,trim={1cm 0.4cm 1cm 0.1cm},,clip]{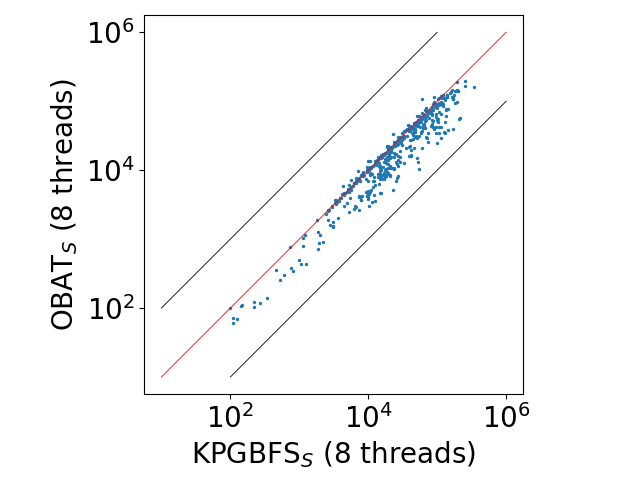}
    \caption{}
  \end{subfigure}
  \begin{subfigure}[]{0.18\columnwidth}
    \includegraphics[width=\textwidth,trim={1cm 0.4cm 1cm 0.1cm},,clip]{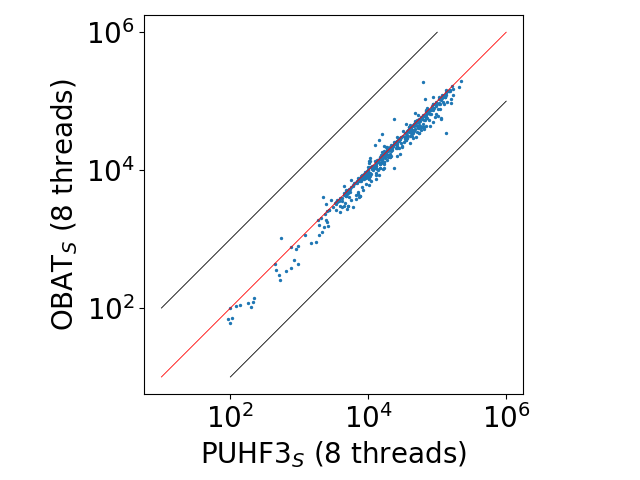}
    \caption{}
  \end{subfigure}
  \begin{subfigure}[]{0.18\columnwidth}
    \includegraphics[width=\textwidth,trim={1cm 0.4cm 1cm 0.1cm},,clip]{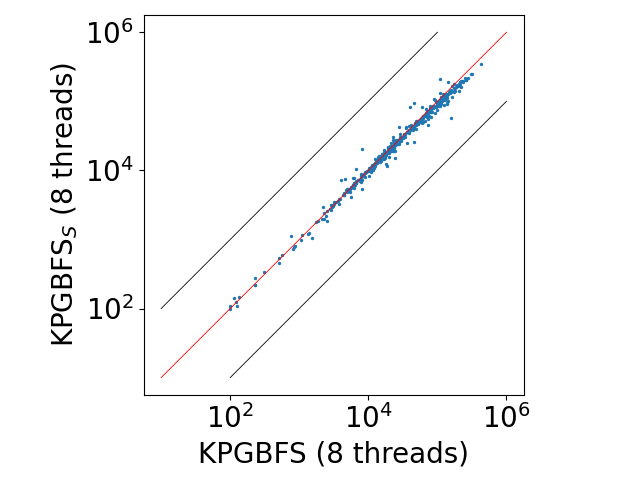}
    \caption{}
  \end{subfigure}
  \begin{subfigure}[]{0.18\columnwidth}
    \includegraphics[width=\textwidth,trim={1cm 0.4cm 1cm 0.1cm},,clip]{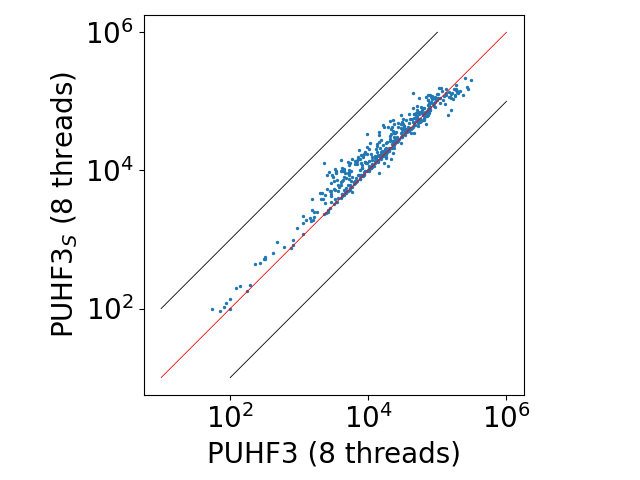}
    \caption{}
  \end{subfigure}

  \begin{subfigure}[]{0.18\columnwidth}
    \includegraphics[width=\textwidth,trim={1cm 0.4cm 1cm 0.1cm},,clip]{figures/evaluation_rate/16threads/evaluation_rate_OBAT_S_K=16_vs_OBAT_K=16.png}
    \caption{}
  \end{subfigure}
  \begin{subfigure}[]{0.18\columnwidth}
    \includegraphics[width=\textwidth,trim={1cm 0.4cm 1cm 0.1cm},,clip]{figures/evaluation_rate/16threads/evaluation_rate_OBAT_S_K=16_vs_KPGBFS_S_K=16.png}
    \caption{}
  \end{subfigure}
  \begin{subfigure}[]{0.18\columnwidth}
    \includegraphics[width=\textwidth,trim={1cm 0.4cm 1cm 0.1cm},,clip]{figures/evaluation_rate/16threads/evaluation_rate_OBAT_S_K=16_vs_PUHF3_S_K=16.png}
    \caption{}
  \end{subfigure}
  \begin{subfigure}[]{0.18\columnwidth}
    \includegraphics[width=\textwidth,trim={1cm 0.4cm 1cm 0.1cm},,clip]{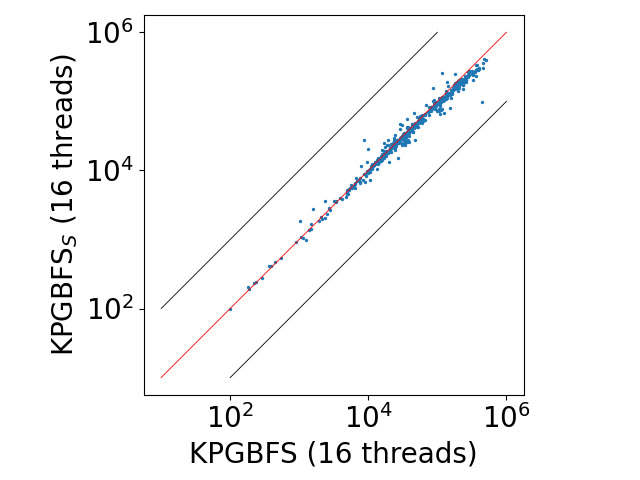}
    \caption{}
  \end{subfigure}
  \begin{subfigure}[]{0.18\columnwidth}
    \includegraphics[width=\textwidth,trim={1cm 0.4cm 1cm 0.1cm},,clip]{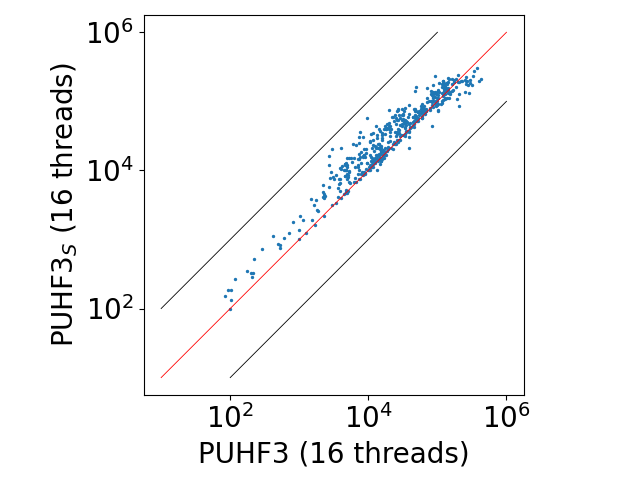}
    \caption{}
  \end{subfigure}

  \caption{ State evaluation rate comparison (states/second), Diagonal lines are $y=0.1x$, $y=x$, and $y=10x$}
  \label{supp:fig:evaluation-rate-comparisons-OBAT_S}

\centering
\end{figure}

\begin{figure}[H]
  \centering
  \begin{subfigure}[]{0.18\columnwidth}
    \includegraphics[width=\textwidth,trim={1cm 0.4cm 1cm 0.1cm},,clip]{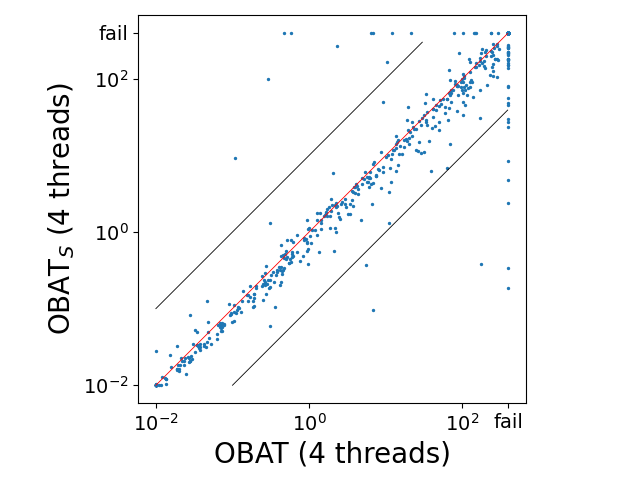}
    \caption{}
  \end{subfigure}
  \begin{subfigure}[]{0.18\columnwidth}
    \includegraphics[width=\textwidth,trim={1cm 0.4cm 1cm 0.1cm},,clip]{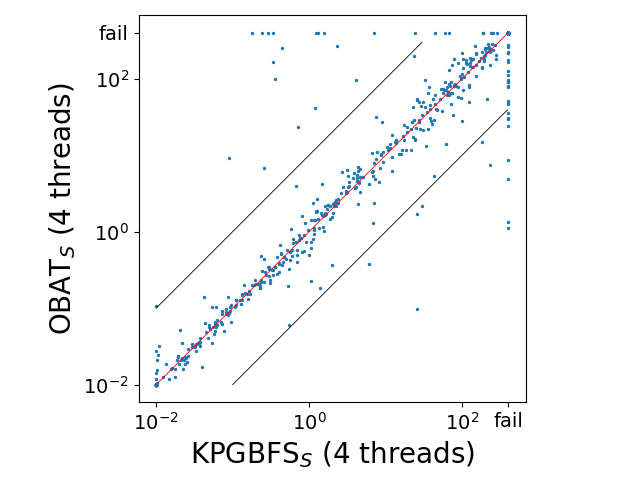}
    \caption{}
  \end{subfigure}
  \begin{subfigure}[]{0.18\columnwidth}
    \includegraphics[width=\textwidth,trim={1cm 0.4cm 1cm 0.1cm},,clip]{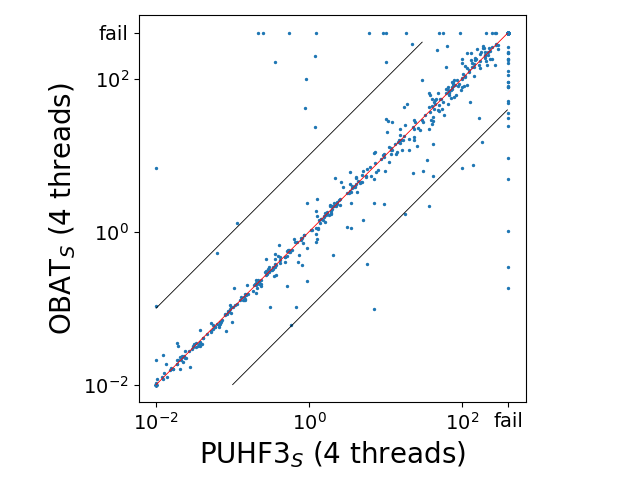}
    \caption{}
  \end{subfigure}
  \begin{subfigure}[]{0.18\columnwidth}
    \includegraphics[width=\textwidth,trim={1cm 0.4cm 1cm 0.1cm},,clip]{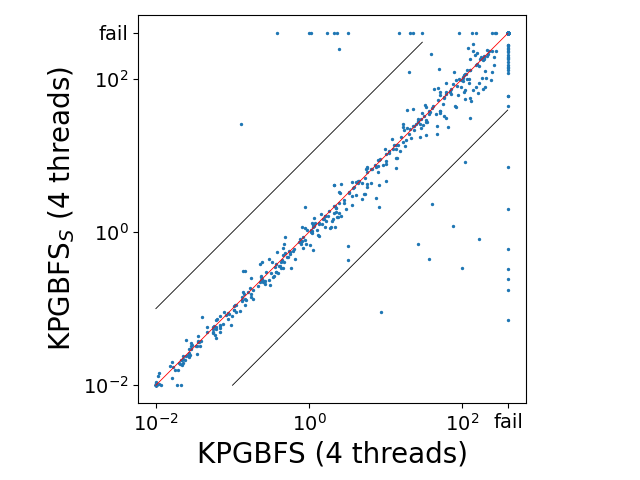}
    \caption{}
  \end{subfigure}
  \begin{subfigure}[]{0.18\columnwidth}
    \includegraphics[width=\textwidth,trim={1cm 0.4cm 1cm 0.1cm},,clip]{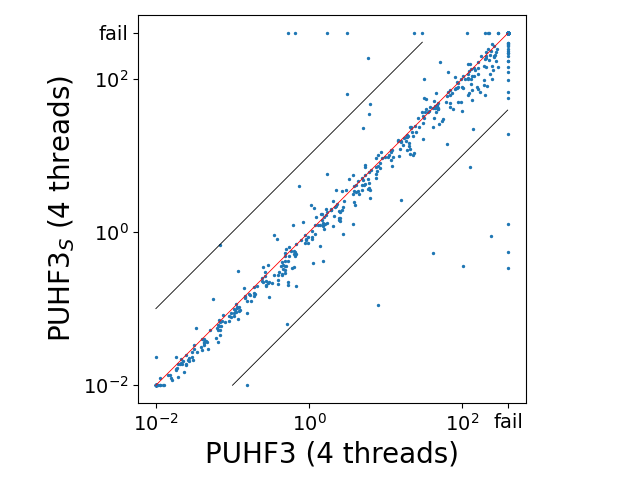}
    \caption{}
  \end{subfigure}

  \begin{subfigure}[]{0.18\columnwidth}
    \includegraphics[width=\textwidth,trim={1cm 0.4cm 1cm 0.1cm},,clip]{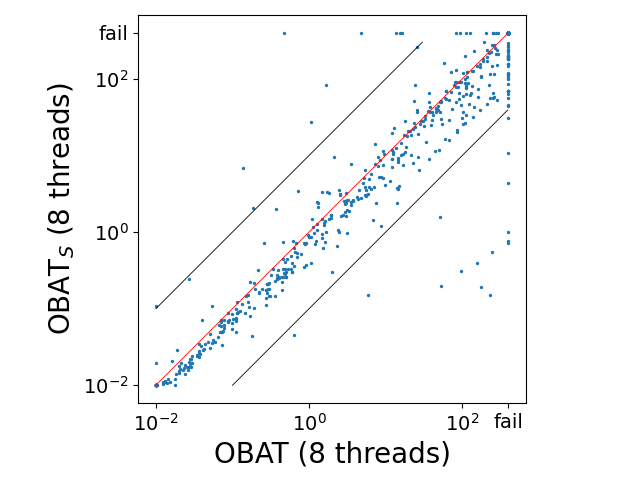}
    \caption{}
  \end{subfigure}
  \begin{subfigure}[]{0.18\columnwidth}
    \includegraphics[width=\textwidth,trim={1cm 0.4cm 1cm 0.1cm},,clip]{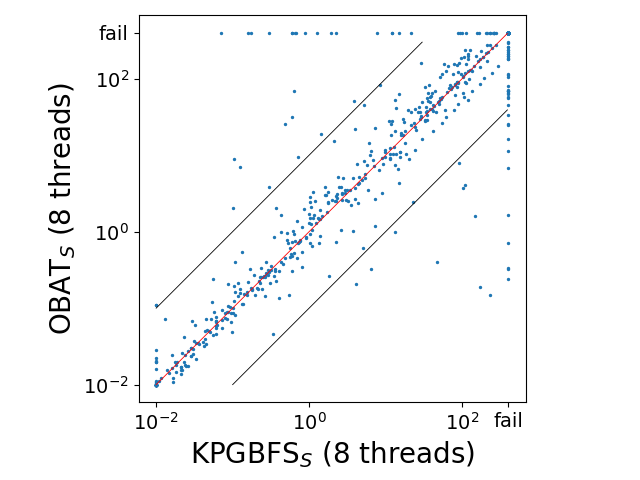}
    \caption{}
  \end{subfigure}
  \begin{subfigure}[]{0.18\columnwidth}
    \includegraphics[width=\textwidth,trim={1cm 0.4cm 1cm 0.1cm},,clip]{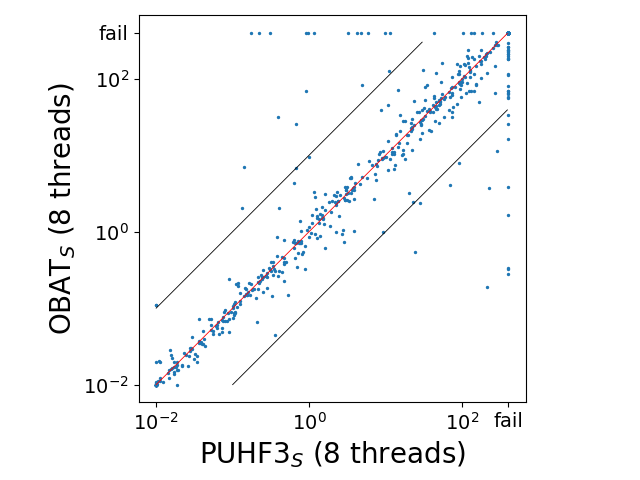}
    \caption{}
  \end{subfigure}
  \begin{subfigure}[]{0.18\columnwidth}
    \includegraphics[width=\textwidth,trim={1cm 0.4cm 1cm 0.1cm},,clip]{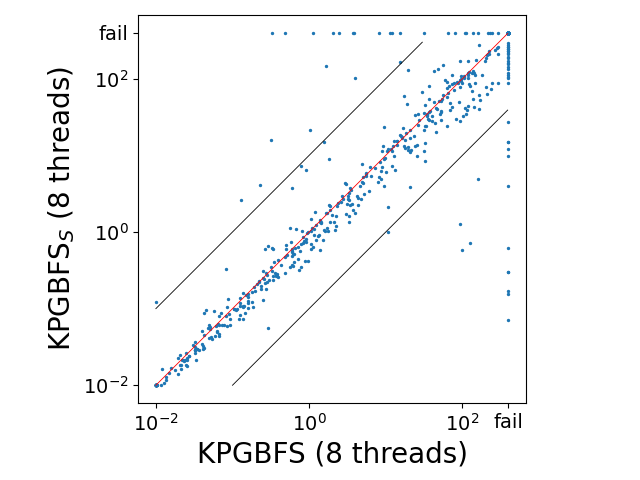}
    \caption{}
  \end{subfigure}
  \begin{subfigure}[]{0.18\columnwidth}
    \includegraphics[width=\textwidth,trim={1cm 0.4cm 1cm 0.1cm},,clip]{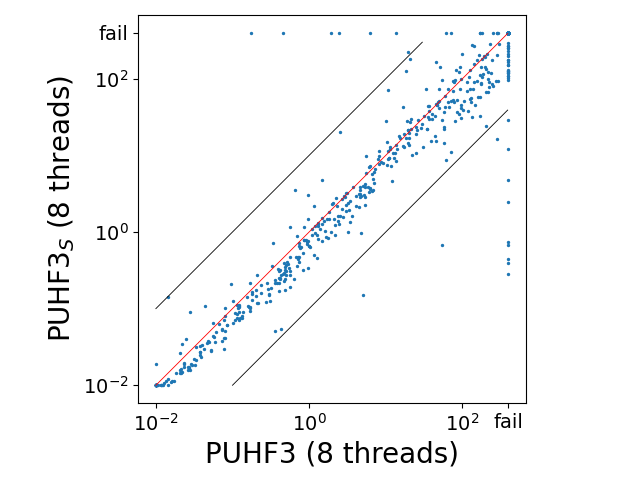}
    \caption{}
  \end{subfigure}

  \begin{subfigure}[]{0.18\columnwidth}
    \includegraphics[width=\textwidth,trim={1cm 0.4cm 1cm 0.1cm},,clip]{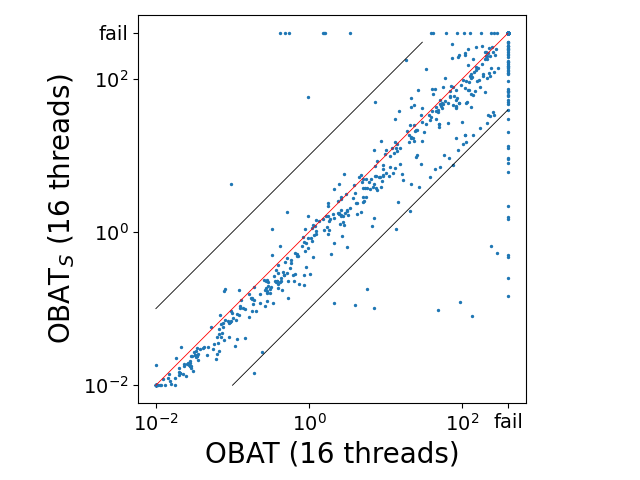}
    \caption{}
  \end{subfigure}
  \begin{subfigure}[]{0.18\columnwidth}
    \includegraphics[width=\textwidth,trim={1cm 0.4cm 1cm 0.1cm},,clip]{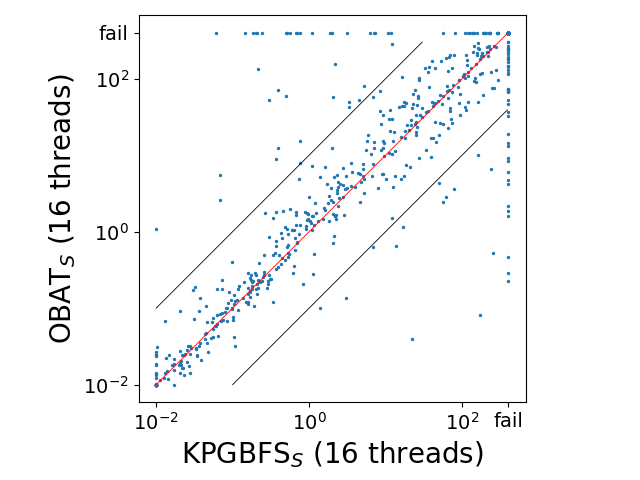}
    \caption{}
  \end{subfigure}
  \begin{subfigure}[]{0.18\columnwidth}
    \includegraphics[width=\textwidth,trim={1cm 0.4cm 1cm 0.1cm},,clip]{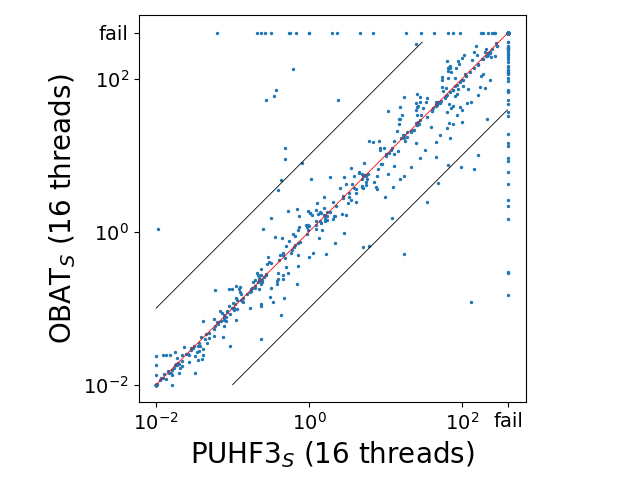}
    \caption{}
  \end{subfigure}
  \begin{subfigure}[]{0.18\columnwidth}
    \includegraphics[width=\textwidth,trim={1cm 0.4cm 1cm 0.1cm},,clip]{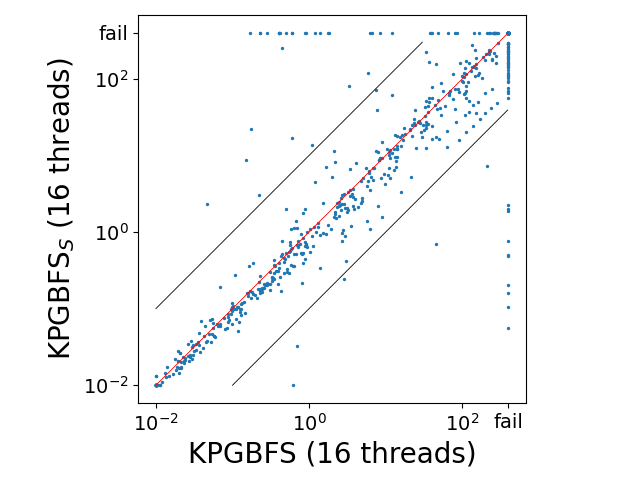}
    \caption{}
  \end{subfigure}
  \begin{subfigure}[]{0.18\columnwidth}
    \includegraphics[width=\textwidth,trim={1cm 0.4cm 1cm 0.1cm},,clip]{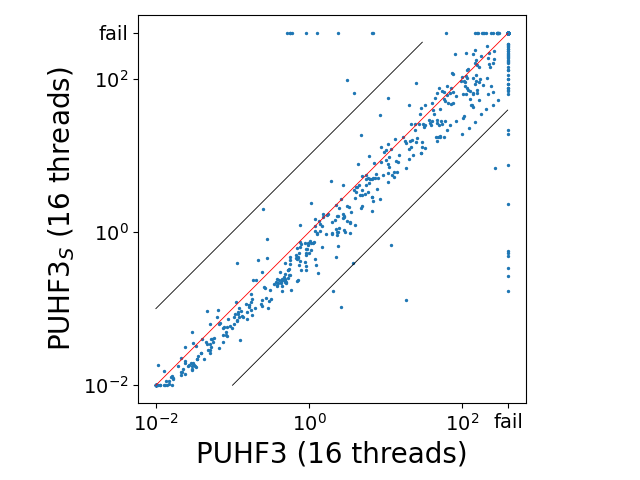}
    \caption{}
  \end{subfigure}
  \centering
  \caption{ Search time (seconds) ``fail''= out of time/memory, diagonal lines are $y=0.1x$, $y=x$, and $y=10x$}
  \label{supp:fig:search-time-comparisons-OBAT_S}
  \centering
\end{figure}

\section{The performance impact of SGE on KPGBFS}

Our experimental results show that \KPDGBFS (KPGBFS with SGE) had higher overall coverage (472/500/532 for 4/8/16 threads) than KPGBFS (462/488/529 for 4/8/16 threads).

In contrast, \cite{ShimodaF24}, the experimental evaluation of SGE on KPGBFS showed that
\KPDGBFS had slightly lower total coverage (507/529/565 for 4/8/16 threads) than KPGBFS (510/534/567 for 4/8/16 threads).

These results are {\it not} contradictory -- the difference is due to the fact that Shimoda and Fukunaga (\citeyear{ShimodaF24}) used the unmodified version of the AutoScale-21.11 benchmark set, while we used a modified version of the AutoScale-21.11 benchmark set.
As explained in Section \ref{sec:bepuhf-experiment} and Supplement Section \ref{supp:benchmark selection}, we replaced
\pddl{gripper} and \pddl{miconic} with harder instances as all 30 of the original instances were solved by all methods and therefore unsuitable (too easy)  for a performance evaluation.

If we remove the new \pddl{gripper} and \pddl{miconic} instances from the total coverage, and then replace them with the much easier, original, AutoScale-21.11 \pddl{gripper} and \pddl{miconic} instances used by \cite{ShimodaF24}, the total coverage for \KPDGBFS is 509/530/556 for 4/8/16 threads,
and total coverage for KPGBFS is 509/532/573 for 4/8/16 threads,
which is similar to the results in \cite{ShimodaF24}.

%% file: paper-aaai25-arxiv.bbl
\begin{thebibliography}{15}
\providecommand{\natexlab}[1]{#1}

\bibitem[{Burns et~al.(2010)Burns, Lemons, Ruml, and Zhou}]{BurnsLRZ10}
Burns, E.; Lemons, S.; Ruml, W.; and Zhou, R. 2010.
\newblock Best-First Heuristic Search for Multicore Machines.
\newblock \emph{{J. Artif. Intell. Res.}}, 39: 689--743.

\bibitem[{Doran and Michie(1966)}]{DoranM66}
Doran, J.; and Michie, D. 1966.
\newblock Experiments with the Graph Traverser Program.
\newblock In \emph{Proc. Royal Society A: Mathematical, Physical and
  Engineering Sciences}, volume 294, 235--259.

\bibitem[{Hart, Nilsson, and Raphael(1968)}]{HartNR68}
Hart, P.~E.; Nilsson, N.~J.; and Raphael, B. 1968.
\newblock {A Formal Basis for the Heuristic Determination of Minimum Cost
  Paths}.
\newblock \emph{{IEEE} Trans. on Systems Science and Cybernetics}, 4(2):
  100--107.

\bibitem[{Heusner, Keller, and Helmert(2017)}]{HeusnerKH17}
Heusner, M.; Keller, T.; and Helmert, M. 2017.
\newblock Understanding the Search Behaviour of Greedy Best-First Search.
\newblock In \emph{{Proc. SOCS}}, 47--55.

\bibitem[{Heusner, Keller, and Helmert(2018)}]{HeusnerKH18}
Heusner, M.; Keller, T.; and Helmert, M. 2018.
\newblock Best-Case and Worst-Case Behavior of Greedy Best-First Search.
\newblock In \emph{{Proc. IJCAI}}, 1463--1470.

\bibitem[{Hoffmann and Nebel(2001)}]{Hoffmann01}
Hoffmann, J.; and Nebel, B. 2001.
\newblock {The FF Planning System: Fast Plan Generation through Heuristic
  Search}.
\newblock \emph{{J. Artif. Intell. Res.}}, 14: 253--302.

\bibitem[{Kishimoto, Fukunaga, and Botea(2013)}]{KishimotoFB13}
Kishimoto, A.; Fukunaga, A.; and Botea, A. 2013.
\newblock Evaluation of a simple, scalable, parallel best-first search
  strategy.
\newblock \emph{{Artif. Intell.}}, 195: 222--248.

\bibitem[{Kuroiwa and Fukunaga(2019)}]{KuroiwaF2019}
Kuroiwa, R.; and Fukunaga, A. 2019.
\newblock On the Pathological Search Behavior of Distributed Greedy Best-First
  Search.
\newblock In \emph{{Proc. ICAPS}}, 255--263.

\bibitem[{Kuroiwa and Fukunaga(2020)}]{KuroiwaF2020}
Kuroiwa, R.; and Fukunaga, A. 2020.
\newblock Analyzing and Avoiding Pathological Behavior in Parallel Best-First
  Search.
\newblock In \emph{{Proc. ICAPS}}, 175--183.

\bibitem[{Phillips, Likhachev, and Koenig(2014)}]{PhillipsLK14}
Phillips, M.; Likhachev, M.; and Koenig, S. 2014.
\newblock PA*SE: Parallel A* for Slow Expansions.
\newblock In \emph{{Proc. ICAPS}}, 208--216.

\bibitem[{Shimoda and Fukunaga(2023)}]{ShimodaF23}
Shimoda, T.; and Fukunaga, A. 2023.
\newblock Improved Exploration of the Bench Transition System in Parallel
  Greedy Best First Search.
\newblock In \emph{{Proc. SOCS}}, 74--82.

\bibitem[{Shimoda and Fukunaga(2024)}]{ShimodaF24}
Shimoda, T.; and Fukunaga, A. 2024.
\newblock Separate Generation \linebreak and Evaluation for Parallel Greedy
  Best-First Search.
\newblock In \emph{Proceedings of ICAPS 2024 Workshop on Heuristics and Search
  for Domain-Independent Planning (HSDIP)}.

\bibitem[{Torralba, Seipp, and Sievers(2021)}]{torralba-et-al-icaps2021}
Torralba, {\'A}.; Seipp, J.; and Sievers, S. 2021.
\newblock Automatic Instance Generation for Classical Planning.
\newblock In \emph{{Proc. ICAPS}}, 376--384.

\bibitem[{Vidal, Bordeaux, and Hamadi(2010)}]{VidalBH10}
Vidal, V.; Bordeaux, L.; and Hamadi, Y. 2010.
\newblock Adaptive K-Parallel Best-First Search: {A} Simple but Efficient
  Algorithm for Multi-Core Domain-Independent Planning.
\newblock In \emph{{Proc. SOCS}}, 100--107.

\bibitem[{Wilt and Ruml(2014)}]{WiltR14}
Wilt, C.~M.; and Ruml, W. 2014.
\newblock Speedy Versus Greedy Search.
\newblock In \emph{{Proc. SOCS}}, 184--192.

\end{thebibliography}
